# Deep HST/STIS Visible-Light Imaging of Debris Systems around Solar Analog Hosts

Short Title: STIS IMAGING OF SOLAR ANALOG DEBRIS SYSTEMS

Glenn Schneider

Steward Observatory and the Department of Astronomy, The University of Arizona

933 North Cherry Avenue, Tucson, AZ 85721 USA

gschneider@as.arizona.edu

Carol A. Grady

Eureka Scientific

2452 Delmer, Suite 100, Oakland, CA 96002 USA

Christopher C. Stark

NASA/Goddard Space Flight Center, Exoplanets & Stellar Astrophysics Laboratory

Code 667, Greenbelt, MD 20771

Andras Gaspar

Steward Observatory and the Department of Astronomy, The University of Arizona

933 North Cherry Avenue, Tucson, AZ 85721 USA

Joseph Carson

Department of Physics and Astronomy, College of Charleston

66 George Street, Charleston, SC 29424

John H. Debes

Space Telescope Science Institute

3700 San Martin Drive, Baltimore, MD 21218 USA

Thomas Henning

Max-Planck-Institut für Astronomie

Königstuhl 17, 69117, Heidelberg, Germany

Dean C. Hines

Space Telescope Science Institute

3700 San Martin Drive, Baltimore, MD 21218 USA

Hannah Jang-Condell

Department of Physics and Astronomy, University of Wyoming

Laramie, WY 82071, USA

Marc J. Kuchner

NASA/Goddard Space Flight Center, Exoplanets & Stellar Astrophysics Laboratory

Code 667, Greenbelt, MD 20771

Marshall Perrin

Space Telescope Science Institute

3700 San Martin Drive, Baltimore, MD 21218 USA

Timothy J. Rodigas

Department of Terrestrial Magnetism, Carnegie Institute of Washington

5241 Branch Road, NW, Washington DC 20015 USA

Motohide Tamura

The University of Tokyo, National Astronomical Observatory of Japan

2-21-1 Osawa, Mitaka, Tokyo 181-8588 JAPAN

John P. Wisniewski

H. L. Dodge Department of Physics and Astronomy, University of Oklahoma

440 West Brooks Street, Norman, OK 73019 USA

# ABSTRACT

We present new Hubble Space Telescope observations of three *a priori* known starlight-scattering circumstellar debris systems (CDSs) viewed at intermediate inclinations around nearby close-solar analog stars: HD 207129, HD 202628, and HD 202917. Each of these CDSs possesses ring-like components that are more-massive analogs of our solar system's Edgeworth-Kuiper belt. These systems were chosen for follow-up observations to provide higher-fidelity and better sensitivity imaging for the sparse sample of solar-analog CDSs that range over two decades in systemic ages with HD 202628 and HD 202917 (both ~ 2.3 Gyr) currently the oldest CDSs imaged in visible or near-IR light. These deep (10 - 14 ksec) observations, with six-roll point-spread-function template subtracted visible-light coronagraphy using the Space Telescope Imaging Spectrograph, were designed to better reveal their angularly large, diffuse/low surface brightness, debris rings, and for all targets probe their exo-ring environments for starlight-scattering materials that present observational challenges for current ground-based facilities and instruments. Contemporaneously also observing with a narrower occulter position, these observations additionally probe the CDS endo-ring environments seen to be relatively devoid of scatterers. We discuss the morphological, geometrical, and photometric properties of these CDSs also in the context of other FGK-star hosted CDSs we have previously imaged as a homogeneously observed ensemble. From this combined sample we report a general decay in quiescent disk $F_{disk}/F_{star}$ optical brightness ~ $t^{-0.8}$, similar to what is seen in at thermal IR wavelengths, and CDSs with a significant diversity in scattering phase asymmetries, and spatial distributions of their starlight-scattering grains.

# 1. INTRODUCTION

Circumstellar (CS) debris systems (CDSs; a.k.a. "disks") are increasingly observed to possess ring-like components or architectures that have been predicted from both thermal infra-red excesses prior to direct imaging and from theoretical models of disk dynamics. Such structures may be indicative of perturbing forces from unseen co-orbiting planets, or other influences (e.g., in the presence of remnant gas) that may sculpt the starlight-scattering materials in these systems into ring-like morphologies. High contrast imaging of circumstellar CDSs at optical and near-IR wavelength with space-based coronagraphy, i.e., with the Hubble Space Telescope (*HST*), and ground-based adaptive-optics augmented imaging systems on large telescopes (e.g., GPI, SPHERE, MagAO, SCExAO and others) have recently informed a rich diversity in CDS architectures and properties for the small number of systems currently within observational reach. This includes an oft-occurrence of starlight-scattering debris rings when observable and exo-ring material that may be bound to, or escaping from, the CDSs (e.g., Schneider et al. 2014; henceforth Sch14).

This paper presents the principle observational results from a high-fidelity CDS optical imaging program using the *HST* Imaging Spectrograph (STIS) undertaken, in part, to study the ring-like debris systems hosted by three nearby close-solar (G2V - G7V) analog stars. Herein we focus on the observational data themselves with some relevant discussions of interpretive analysis. More in depth modeling and further consideration of the analysis quality scattered-light images presented within are the subjects of future papers.

In § 1 of this paper we introduce the continuing relevance and utility of the symbiotic uniqueness space for *HST*/STIS scattered-light imaging of CDS materials and architectures. In § 2 we discuss and present the target sample in the context also of other related and precursor observations. In § 3 we discuss the observational plan and technical methodology. In § 4 we review the need for, and our selection of, point spread function (PSF) calibration template stars. In § 5 we review the data reduction and calibration processes and techniques applied to the observational data. In § 6 we describe a simple morphological scattered-light model for ring-like components of CDSs that we have employed to estimate their primary characterizing geometrical and physical parameters. In § 7 we discuss in detail the individual objects: HD 207129, HD 202628, and HD 202917. In § 8 we discuss and intercompare the observationally derived results as a target class in the context of, contributing to, and expanding the prior sparse sample of imaged solar-analog CDSs. In § 9 we provide some closing commentary.

## 2. THE *HST*/GO 13786 TARGET SAMPLE

In 2010 – 2011, in *HST* /GO program 12228[1], we observed an *HST*-selected sample of 11 CDSs comprised of 10 debris disks and one mature protoplanetray disk. These CDSs sampled a diverse set of main-sequence host stars ranging from M1V - A5V in spectral type and ~ 10 – 1000 Myr in age; see Sch14 and in context of Choquet 2016 (*c.f.*, their Fig 1). These were high-fidelity follow-ups to prior discovery observations from *HST*'s Near Infrared Camera and Multi-Object Spectrometer (NICMOS) and Advanced Camera for Surveys (ACS) instruments, homogeneously obtained using *HST* 's Imaging Spectrograph (STIS) with 6-roll PSF template subtracted coronagraphy (6R-PSFTSC) for all targets. This sample included a number of intermediate to high inclination CDSs from which ring-like debris components were directly observed. This included HD 181327, HD 61005, HD 15115, HD 92945, and the close-solar (G2V) analog HD 107146[2] . Most of these, as well as the nearly edge-on HD 32297, were also shown to possess extensive systems of exo-ring materials extending well beyond their bright ring radii, posited as possibly commonly occurring in CDSs. With this initial sample in hand, in follow-on with *HST* GO program 13786, we sought to extend in number and "fill-in" in age and spectral type this observationally homogeneous, but small, sample with a focused target list of five nearby intermediate inclination CDSs each with *a priori* established ring-like morphologies.

---
[1] http://www.stsci.edu/cgi-bin/get-proposal-info?id=12228&observatory=HST
[2] Herein, in § 8.2.1, we also call attention also to the directly-observed ring-like component in HD 15745.

With sufficient spatial resolution and contrast[3] in scattered-light imaging, many CDSs are directly seen to possess ring-like architectures or components that may be indicative of planet presence. These are best revealed with spatially resolved imaging in systems with low to intermediate inclinations from the line of sight; i.e., close to "face on". Although such low-inclination systems provide revealing views of the radial distribution of starlight-scattering material (e.g., Debes et al. 2016), they leave interpretive analysis of particle properties augmented through scattering phase functions ill-constrained[4] (Stark et al. 2014, Hedman & Stark 2015). Additionally, scattered light images of low-inclination CDSs, while enabling best spatially-resolved 2D mapping of the in-plane scattering material (e.g., HD 181327; Stark et al. 2015), provide limited information on the vertical structure in such systems. Conversely, azimuthal structure in edge-on (high-inclination) disks such as β Pic (Apai et al. 2015) and AU Mic (Sch14, Boccaletti et al. 2015) is self-obscured by the viewing geometry, though with better views of the vertical (out of disk-plane) material in the system. While high-inclination CDSs provide nearly a full range of scattering phase angles, the tomographic view of the scattering particles is conflated by the line-of-sight integrated viewing geometry through the plane of the disk. Therefore, intermediate inclination CDSs can provide a best "compromise" in viewing geometries to simultaneously explore the in-plane and vertical distribution of starlight-scattering

---

[3] In this paper, image contrast is defined as the ratio between the flux density in a resel at any stellocentric location after coronagraphic and/or augmented starlight suppression by any other means including point-spread-function (PSF) subtraction, compared to the flux density in the central resel of an unocculted stellar PSF.
[4] explorable scattering phase angle range: $\varphi = \cos^{-1}(\sin i \cos\theta)$; $i$ = inclination, $\theta$ = azimuth angle in disk plane

materials and provide largely unobstructed views of the system architectures. Many of these CDS structures currently remain best revealed with *HST* coronagraphy symbiotic to capabilities provided by suitably instrumented large ground-based telescopes with high-order adaptive-optics systems.

Our GO 13786 sample included two, *a priori* only sparsely-sampled (with direct imaging), target classes: First, and the subject of this paper, a G-star sample of three close-solar analog stars both younger (<~ 80 My) and older (>~ 2 Gyr) than those imaged prior in *HST*/GO 12228. The principal characteristics of these targets are summarized in Table 1. Second, a "revisit" to better explore both the innermost, and the outermost, light-scattering regions of two iconic and youthful (~ 5 - 8 Myr) higher-stellar mass (spectral types A0V, B9V) debris systems: HR 4796A and HD 141569A - both observationally challenging by the presence of M-star companions. These are the subjects of two companion papers, Schneider et al. 2016 (in prep.) and Konishi et al. 2016, as discussed further therein.

**Table 1**
HST GO 13786 G-star Target Sample

| Target | Vmag | B-V | Spec | Dist. (pc) | Age[a] (Myr) | Disk $L_{IR}/L_{star}$[b] | Initial *HST* Imaging | |
|---|---|---|---|---|---|---|---|---|
| | | | | | | | Instrument | Reference |
| HD 207129 | 5.58 | +0.60 | G2V | 16.0 | 2350 ± 850 | 1.1 x $10^{-4}$ | ACS | Krist et al. 2010 |
| HD 202628 | 6.75 | +0.64 | G5V | 24.4 | 2300 ± 1000 | 1.4 x$10^{-4}$ | ACS | Krist et al. 2012 |
| HD 202917 | 8.67 | +0.65 | G7V | 43.0 | 30 (+10, - 20) | 2.9x$10^{-4}$ | NICMOS | Soummer et al. 2014 |

[a] Age estimations –   HD 202917: Moór et al. 2006; HD 207129: Mamajek & Hillenbrand 2008, Soderbloom 1985, HD 202628: Krist et al. 2012
[b] $L_{IR}/L_{star}$ estimations –  HD 202917: Moór et al. 2006; HD 207129: Marshall et al. 2011, HD 202628: Koerner et al. 2010

### 3. OBSERVATIONS

We obtained STIS coronagraphic observations of our three G-star CDS targets, along with their contemporaneously interleaved PSF-template stars, as part of *HST*/GO program 13786[5] (G. Schneider: PI). All observations executed during the interval 2015 May – October. For all targets we closely followed the successful observational paradigm for *HST* multi-roll PSF-template subtraction coronagraphy as previously demonstrated in *HST* GO program 12228 as detailed by Sch14. For each G-star CDS and its paired PSF template star, two sets of four contiguous single-orbit visits were programed for coronagraphic observations using STIS occulting wedge A. In each orbit, Wedge A was used at both its 0.6" (WA0.6) and 1.0" (WA1.0) wide locations for, in combination with exposure times used (see Table 2), high-dynamic range imaging (with instrumental GAIN=4) over stellocentric angles ranging from 0.3" to > 10". Each CDS target was observed at six spacecraft roll angles in two sets of three orbits separated by 2 – 4 months due to *HST* differential roll constraints at any epoch[6]. At each epoch the PSF template star for each CDS target was similarly observed in a single orbit interleaved, contiguously, between the second and third CDS orbits (thus a total of 8 orbits consumed for each CDS target and its PSF template star). Within each orbit, the target (CDS-host or PSF-template star) was first placed at the WA0.6 position, imaged for roughly 1/3 of an orbit multiple times to a nearly full-well depth where brightest (not to exceed image saturation at the edges of the occulting wedge), then repositioned at WA1.0 and more deeply imaged multiple times for the remainder of the orbit.

---
[5] http://www.stsci.edu/cgi-bin/get-proposal-info?id=13786&observatory=HST

Table 2 gives a summary of the observations, with further details available through the STScI MAST archive. The rationale for the details of the observational design are given in Sch14. Columns 1 and 2, respectively, give the disk-host and corresponding interleaved PSF template star names, and the UTC date of the start of each of the 4-orbit contiguous observing sequences. Column 3 gives the absolute orientation angle of the images (FITS header keyword ORIENTAT) measured from the image +Y axis to celestial north counter-clockwise through east about the target star. Column 4 indicates the total number of individual WA0.6 exposures for that epoch, equally divided with multiple visits. E.g., for the first entry in the table (HD 201917) is 3 visits x 8 exposures per visit = 24 exposures. Column 5 gives the total exposure time for all corresponding WA0.6 exposures in the visit set. E.g., 24 exposures x 100 s per exposure = 2400 s. Columns 6 and 7 are similar to Columns 4 and 5, but for WA1.0 images. Column 8 gives the visit number (ID) as used in the MAST identification of data sets unique to GO program 13786 (ocjc + data ID).

---

[6] The absolute celestial orientation angles for all G-star visits were unconstrained in scheduling the observations, but the relative orientations (differential between visits) for the CDS targets were, optimally (ideally) constrained for sequential visit-to-visit orientation differentials of ~ 30°. This, however, was not always closely realizable due to *HST* guide star and other pointing restrictions; see Table 2 for details of the as-observed "orient" angles.

**Table 2**
Observation and Data Log for *HST* GO 13786 Solar Analog Targets and PSF Template Stars

| Target (Disk/PSF) | UT Date Obs. Start | Orient ° | W0.6 # Exp | W0.6 $T_{EXP}$ *all visits* sec | W1.0 # Exp | W1.0 $T_{EXP}$ *all visits* sec | Visit Data ID[a] |
|---|---|---|---|---|---|---|---|
| HD 202917 | 13OCT2015 | 36.1, 14.1, 352.1 | 24 | 2400.0 | 9 | 4437.0 | 51, 52, 54 |
| LTT 8893 | 13OCT2015 | 2.6 | 10 | 264.0 | 13 | 1677.0 | 53 |
| HD 202917 | 01AUG2015 | 321.1, 300.6, 280.1 | 24 | 2400.0 | 9 | 4797.0 | 55, 56, 58 |
| LTT 8893 | 01AUG2015 | 285.1 | 10 | 264.0 | 4 | 2114.4 | 57 |
| HD 207129 | 26MAY2015 | 236.7, 214.2, 191.7 | 72 | 417.6 | 39 | 4524.0 | 15, 16, 18 |
| Tau$^1$ Gru | 26MAY2015 | 203.1 | 23 | 204.7 | 8 | 1427.2 | 17 |
| Tau$^1$ Gru-cal[b] | 26MAY2015 | 203.1 | --- | --- | 2 | 178.4 | 17 |
| HD 207129 | 08AUG2015 | 311.7, 289.7, 267.7 | 72 | 417.6 | 39 | 4524.0 | 11, 12, 14 |
| Tau$^1$ Gru | 08AUG2015 | 268.3 | 23 | 204.7 | 8 | 1427.2 | 13 |
| Tau$^1$ Gru-cal[b] | 08AUG2015 | 268.3 | --- | --- | 2 | 178.4 | 13 |
| HD 202628 | 17SEP2015 | 341.1, 1.6, 21.1 | 60 | 1032.0 | 15 | 4890 | 01, 02, 04 |
| HR 8042 | 17SEP2015 | 6.5 | 20 | 310.0 | 6 | 1320 | 03 |
| HD 202628 | 30MAY2015 | 243.1, 220.1, 197.1 | 60 | 1032.0 | 15 | 4890 | 05, 06, 08 |
| BX Mic[c] | 30MAY2015 | 215.1 | 20 | 366.0 | 5 | 1570 | 07 |

[a] (Single-orbit) visit level dataset ID as assigned by MAST. GO 13786 data archived as ocjc + datset_id + *.
[b] Same target – Tau$^1$ Gru: Calibration and check exposures - not used directly for PSF subtraction template creation.
[c] V07: Close angular proximity background star and/or companion revealed – ill-suited as PSF subtraction template.

## 4. PSF-TEMPLATE STARS

Beyond raw coronagraphy, the remaining incompletely suppressed starlight in the PSF halo polluting a CDS image is further reduced by subtracting a PSF template derived from a diskless star observed in a nearly identical manner. PSF-template stars of at least roughly comparable brightness to paired CDS stellar host targets were selected and scheduled with temporal, roll-range, and pointing constraints, to minimize otherwise systematically enhanced PSF-subtraction residuals. Template stars were selected and observations were planned with respect to their paired CDS targets to simultaneously meet the following criteria as closely as possible: (a) identically matched in B-V color (Δ[B-V] = 0.0), (b) change in telescope attitude (slew distance) from target to template star and back to target of < 10°, (c) change in telescope roll angle < 20° from target star middle (of three) visits to template star, (d) contiguous 4-orbit visit sets with interleaved PSF visit as third with no interruptions other than Earth occultations. Details for the PSF star selection and scheduling are given in Table 3.

Table 3
Planned, Contemporaneously Interleaved, PSF Template Stars

| Disk Target | PSF Star | PSF Spec | PSF V | PSF B-V | Δ(BV)[a] | PSF Slew[b] | PSF ΔONR[c] | |
|---|---|---|---|---|---|---|---|---|
| HD 207129 | Tau1 Gru | G0V | 6.04 | +0.62 | –0.03 | 11.0° | +11.1° | +21.4° |
| HD202628 | BX Mic[b] | G0IV | 6.82 | +0.64 | 0.00 | 10.4 ° | -25.4° | --- |
|  | HR 8042 | G3IV | 6.64 | +0.68 | –0.04 | 2.3° | --- | +5.0° |
| HD 202917 | LTT 8893 | G3V | 7.22 | +0.66 | –0.01 | 7.7° | +11.5° | +15.5° |

[a] Difference in B-V color index between target and PSF template stars; Δ(VR) used as a secondary ranking criterion.
[c] ONR: "Off Nominal Roll"; ΔONR: roll angle change between 2nd target & PSF star visits in contemporaneous set.

## 5. DATA CALIBRATION AND REDUCTION

*Instrumental Calibration and Reduction:* In processing the raw coronagraphic images into fully reduced "analysis quality" (AQ) images for metrical and interpretive analysis, we follow the methodology and techniques discussed by Sch14 to which the reader is referred for details. In summary: (1) We instrumentally calibrate the individual raw images to produce bias, dark current, non-linearity, and flat-field corrected images (FLT files) using the *calstis* S/W with calibration reference files provided by STScI (CDBS system) updated to the epoch of the individual observations. (2) The location of the occulted target star in each FLT image (and its uncertainty by dispersion) is determined using the "X marks the spot method", described by Sch14 (§ 5.4 therein). (3) For each visit, the multiple FLT images obtained at each occulting wedge location were vetted for anomalies (e.g., image "drift"; none found within typical two-star fine-lock intra-orbit pointing stability) and then median combined (for cosmic ray elimination and building SNR) into visit-level count-rate images. (4) For each visit, the visit-level WA0.6 and WA1.0 PSF template images were subtracted from the corresponding, step 3, target images. Subtraction is accomplished per Sch14 treating the PSF template image intensity (brightness in instrumental counts s$^{-1}$ pixel$^{-1}$) and star location imaged on the detector as free parameters. Differences are simultaneously minimized in iterative subtraction in regions along the *HST* diffraction spikes unaffected by the STIS occulting masks and where not dominated by disk flux in target images per Sch14. (5) Visit-level PSF subtracted images (in the instrument orientation frame) are rotated to a common celestial orientation (north "up") about the occulted star and median combined masking the STIS occulting wedges, diffraction spike residuals, regions of image saturation near the wedge edges, and any image anomalies in individual images. (6) Chromatic residual correction (derived from individual visits) is applied, if needed, for imperfect PSF-template color-matching per Sch14 (*c.f.,* § 5.6 and *e.g.*, Fig.17, therein). (7) The W0.6A and W1.0A images are combined with selective masking for the separate regions unsampled or degraded in one but not the other, producing a "final" (single) AQ data image over the largest stellocentric and azimuth-angle ranges afforded by both occulting wedge locations and very large (target dependent) imaging dynamic range in surface brightness.

*Absolute Photometric Calibration:* In this paper we report photometric observational results in both instrumental count rate units of counts s$^{-1}$ pixel$^{-1}$ (with 1 count = 4.016 photoelectrons for "GAIN=4") and in physical units in flux density (Jy) or surface brightness (Jy arcsec$^{-2}$). For the latter, we adopt the STIS instrumental calibration as codified in the *HST* synthetic photometry package SYNPHOT (calcphot task) with 1 count s$^{-1}$ pixel$^{-1}$ (with GAIN=4) = 4.55 x 10$^{-7}$ Jy, and an absolute photometric zero-point calibration such that a 0 magnitude (Vega system) star produces a total flux density of 2923 Jy in the STIS (unfiltered) 50CCD spectral bandpass. In assessing the brightness of observed CS debris at visible wavelengths, we assume reflectance by the starlight-scattering material is spectrally neutral (i.e., the debris being the same "color" as the host star). This may not be a correct, but no filter-band (or spectro-)photometry necessary to assess a "color correction" under the broad STIS 50CCD bandpass is available (or currently possible), but an assumption of near spectral flatness should be a close approximation. For absolute calibration of surface brightness (SB; i.e., flux density per unit area) we adopt the *a priori* determined image scale of the STIS 50CCD channel as 0.05077 arcsec pixel$^{-1}$ on both square-pixel axes. For SB, this gives 1 count s$^{-1}$ pixel$^{-1}$ = 177 µJy arcsec$^{-2}$ = 18.04 V$_{mag}$ arcsec$^{-2}$.

## 6. SCATTERED-LIGHT IMAGE MODELING

We ascertain estimates of the geometric, astrometric, and photometric characteristics of the debris rings themselves by comparing simple scattering models of the rings to best match their observed SB distributions (e.g., see Schneider et al. 2006). We assume an azimuthally uniform ring with a Gaussian radial surface density distribution of scattering particles and allow for offsets of the ring center from the star. We illuminate models with starlight according to an r$^{-2}$ power law and adopt a Henyey-Greenstein (HG) scattering phase function (1941). We treat as observationally-informed free parameters, initially estimated visually and/or by ellipse fitting: the ring celestial position angle (*P. A.*), inclination (*i*), and apparent semi-major axis (*a*) of an assumed intrinsically circular (but seen in projection) ring ellipse. We also fit the inner and outer radii of the ring as characterized by the half-power points of their respective edge SB profiles, and the slopes of the inner and outer edges. We then additionally vary the 2-dimensional displacement of the ring center from the stellocenter and the HG asymmetry parameter, *g*, through a grid of values with resolution 0.05 in *g*. In least-squares fitting the models to the data, regions both internal and external to the imaged debris rings, where PSF-subtraction and sky-plus-instrumental noise obviously dominates over detectable flux from the ring, were digitally masked. The model free parameters were then iterated to convergence minimizing the variance in the fitting region.

## 7. INDIVIDUAL OBJECTS

### 7.1 – HD 207129

*Introductory Notes.* HD 207129 is a G2V close-solar analog and is one of the two oldest stars (along with HD 202628; § 7.2) for which a CDS has been imaged in scattered-light. The presence of cold dust around HD 207129 was inferred by Walker & Wolsterncroft (1988) from 12.5 – 60 μm IRAS excesses, confirmed by Jordin de Muizon (1999) with 2.5 - 180 μm ISO photometry with far-IR excess in its spectral energy distribution (SED) at > 20 μm. A spatially extended disk was detected in thermal emission in Spitzer/MIPS imaging at 70 μm by Krist et al. (2010). Correcting for the broadening effects of the instrumental beam, they deduced an intrinsic size of an apparently elliptical disk at 70 μm of 18.8" x 8.1" (300 x 130 au). The disk, observed as part of the Herschel/DUNES program, has also been resolved with PACS imaging in thermal emission from 70 – 160 μm (Marshall et al. 2011) as modeled by Lohne et al. (2011). Krist et al. (*ibid*) additionally resolved the disk with *HST*/ACS visible-light coronagraphy at 0.6 μm. The ACS image, derived from two-roll "self-subtraction" of the underlying stellar PSF, revealed a ring-like disk inclined 60° from face-on with a major axis *P. A.* of 127°, in approximate geometrical agreement with the 70 μm MIPS, and Herschel/PACS images.

*Observations and PSF Subtraction (Details).* *HST* /STIS 6R-PSFTSC observations of HD 207129, and its color-matched PSF-template star Tau[1] Gru, were conducted per the GO 13786 observing plan with no variances or anomalies on 2015 May 26 (visits 15 - 18) and 2015 Aug 08 (visits 11 - 14). With visit-level PSF subtraction, a low SB scattered-light excess from the disk was detectable (at low SNR) in each of the individual 1508 s exposure Wedge-A1.0 images, co-rotating with visit-to-visit reorientation of the celestial field. Despite the very close |B-V| color-index matching of its PSF template star, a stellocentric, (first-order) azimuthally-symmetric, and radially-modulated residual pattern (seen also independently in the Wedge-A0.6 images) at r < 6" characteristically betrayed by a small chromatic error in PSF subtraction with stellocentric angle (undersubtraction at r < 1", oversubtraction at 1" ≤ r ≤ 3.5", undersubtraction at 3.5" ≤ r < 6"). We illustrate this (hard stretched for maximum visibility), uncorrected, after combining all images in a common celestial frame in Fig. 1 panel A. To largely mitigate this problem, we follow the empirical color-correction process detailed in Sch14 as applied therein by example for HD 92945 (with source-masking here appropriate for HD 207129); see Fig. 1 panel B. After correction, the WedgeA-1.0 deep images and WedgeA-0.6 shallow images, in combination, both well reveal the ring-like disk reported by Krist et al. (2010) and provide SB limits to any dust (or lack thereof) interior to the starlight-scattering debris ring itself. For better visualization of the ring-like disk, in Fig. 1 panel C we convolve the color-corrected image with a 3x3 pixel boxcar smoothing kernel to reduce the pixel-to-pixel instrumental background noise.

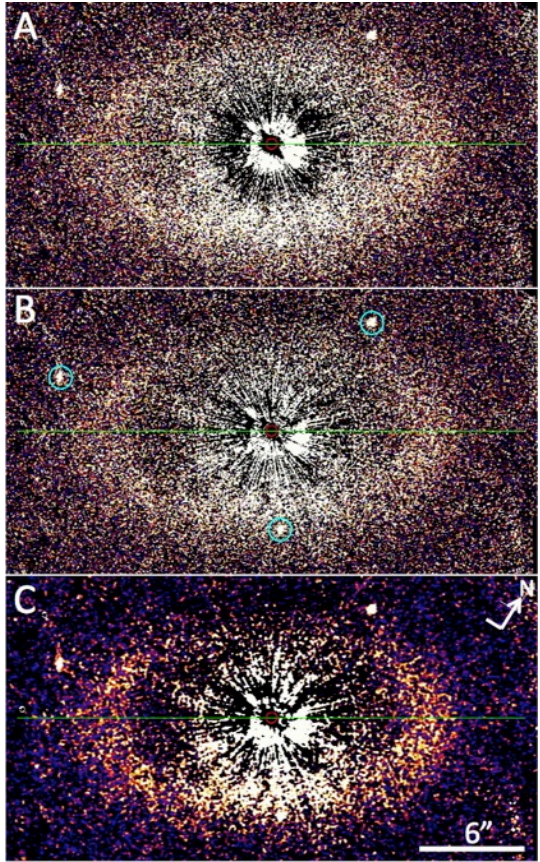

Fig 1. 6R-PSFTSC images of the HD 207129 CDS combining all Wedge-A 0.6 and 1.0 data, rotated 32° clockwise from "north up" to place the morphological disk major axis on the image horizontal. A & B: Respectively, before and after correcting for radially-symmetric chromatic PSF-subtraction residuals. C: Panel B convolved with a 3x3 pixel boxcar smoothing kernel to reduce pixel-to-pixel noise. Three compact/point sources (blue circles in panel B) identified by Krist et al. 2010 are confirmed as background objects through differential proper motions between the ACS epoch 2006 and STIS epoch 2015 images. All panels: FOV = 25.4" x 13.4" and linear display stretch from 0 to +0.015 counts s$^{-1}$ pixel$^{-1}$.

*Principal Results.* The STIS 6R-PSFTSC scattered-light imaging of the HD 207129 CDS (Fig. 1) reveals an angularly large, low SB, ring-like disk at an intermediate inclination angle with a morphology as generally described by Krist et al. 2010. The broad ring-like disk appears to be devoid of starlight scattering material interior to its inner edge to a stellocentric angle limit of r ≈ 5" where, at smaller stellocentric angles, residual starlight dominates the centosymmetric background. The SB ring-model (§ 6) that best fits the observed data (by minimizing the sum of the squares of the observed-minus-model image residuals) is illustrated in Fig. 2, with the model parameters given in Table 4. Although the model also allows for a stellocentric offset, none of significance was evidenced in initial fits to the HD 207129 image, and so was fixed to zero; i.e., constraining the apparent ring center as coincident with the location of the host star. We did not treat $g$ iteratively, but produced a grid of models incrementally increasing from $g = 0.0$ (the isotropic case) in steps of $\Delta g = + 0.05$, for which a best global fit was found for $g \sim 0.25$.

The model appears to give a very good fit overall, though may slightly over-subtract the disk flux asymmetrically w.r.t. the locations (radii) of the peak brightness at some stellocentric azimuth angles. This is marginally evidenced in the somewhat darkest (black) distorted "ring" of low-amplitude correlated negative-going residuals visible in Fig. 2 panel C. This may be indicative of the limitations of the current model that does not account for possible dust density anisotropies or non-HG scattering properties.

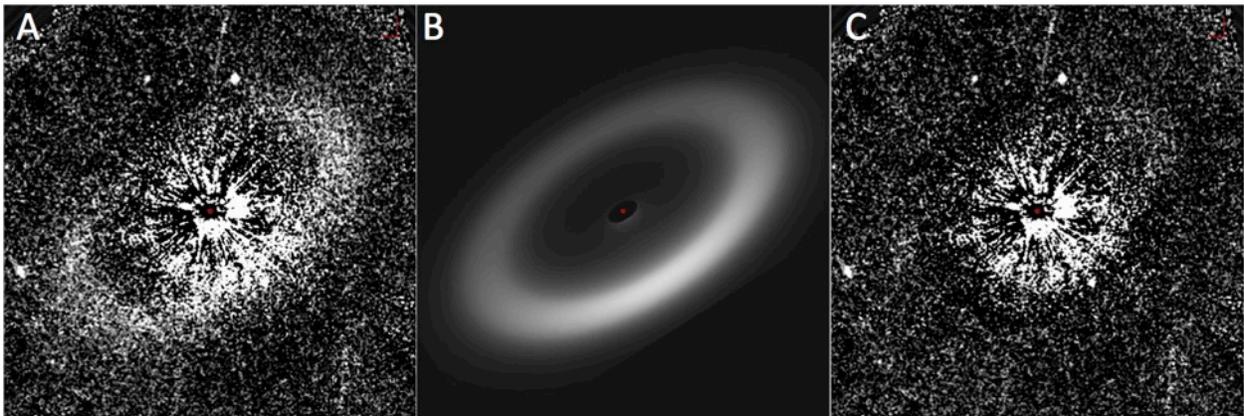

Fig 2. HD 207917 best fit debris ring model. A. Observed image (north up). B. Model with $g = +0.25$ and other parameters as in Table 4. C. Observed-minus-model image residuals. All images: Linear stretch from -0.001 to +0.01 counts s$^{-1}$ pixel$^{-1}$, FOV 512 x 512 pixels (26" x 26") with celestial north up and east left.

**Table 4**
HD 207129 Debris Ring Parameters

| | | |
|---|---|---|
| Inclination (*i*) from face-on | 57.9° ± 2° from face-on | |
| *P. A.* Major Axis | 122° ± 1° E of N | |
| Ring peak SB on major axis (*a*) | 9.34" ± 0.03" | 149.4 ± 1.0 au [a] |
| Ring HWHM major axis inner-edge radius | 7.13" ± 0.04" | 115.7 ± 0.9 au [a] |
| Ring HWHM major axis outer-edge radius | 11.75" ± 0.03" | 188.0 ± 1.2 au [a] |
| Ring width (FWHM; Δr) | 72.3 ± 1.5 au | |
| g (HG scattering asymmetry parameter) | 0.25 ± 0.05 | |
| Total ring flux density[a] | 310 ± 29 c/s/pix | 141 ± 13 μJy |
| Disk (0.6 μm) Scattering Fraction[b] | $8.2 \times 10^{-6} \pm \sim 8 \times 10^{-7}$ | |

[a] includes uncertainty in stellar parallax from SIMBAD ; [b] $V_{star}$ = 5.58 → 17.1 Jy

1. Brightness and Scattering Fraction: In Fig. 3 we compare the observed and model radial SB profiles along the disk major axis where, at the ring ansae, the image contrast is less challenging and more accurately measured than closer to the minor axis. It is in this region that the deficit of light-scattering dust interior to the debris ring itself is visually apparent (e.g., in Fig. 2 panel A.) Using the model parameters in Table 4 closely replicates the main features of the observed profile on both sides of the star across the ring ansae, though some local deviations are seen - in particular here at ~ 5" < r < 7" where the model slightly over-predicts the observed brightness. This could be due to a deficiency in the model, or simply a bias in the signal as the stellocentric angle is approaching the noise dominated regime in the image at r < 5". The peak SB of the debris ring along its major axis, at its diametrically opposed ansae, appears marginally brighter on its northwestern side at ~ 0.96 μJy arcsec$^{-2}$ (23.7 Vmag arcsec$^{-2}$), and 0.68 μJy arcsec$^{-2}$ (24.1 Vmag arcsec$^{-2}$) on its southeastern side. The total 0.6 μm flux density of the disk, as informed by the best-fit model, is 141 ± 13 μJy. This gives rise to a total disk optical scattering fraction, $F_{disk}/F_{star}$, of ~ 8.2 x 10$^{-6}$ to its host star with $V_{mag}$ = +5.58. This is in good agreement with the prior estimation by Krist et al. 2010 of ~ 7.6 x 10$^{-6}$ independently derived with a different data set with iterative roll subtraction. Flux-loss and/or larger photometric uncertainties from "self-subtraction" for some disk geometries with this method are possible. However, the good agreement (≈ ± 5%) in the independently derived global scattering fraction for the HD 207129 CDS suggests the original estimation by Krist et. al 2010 was photometrically well calibrated.

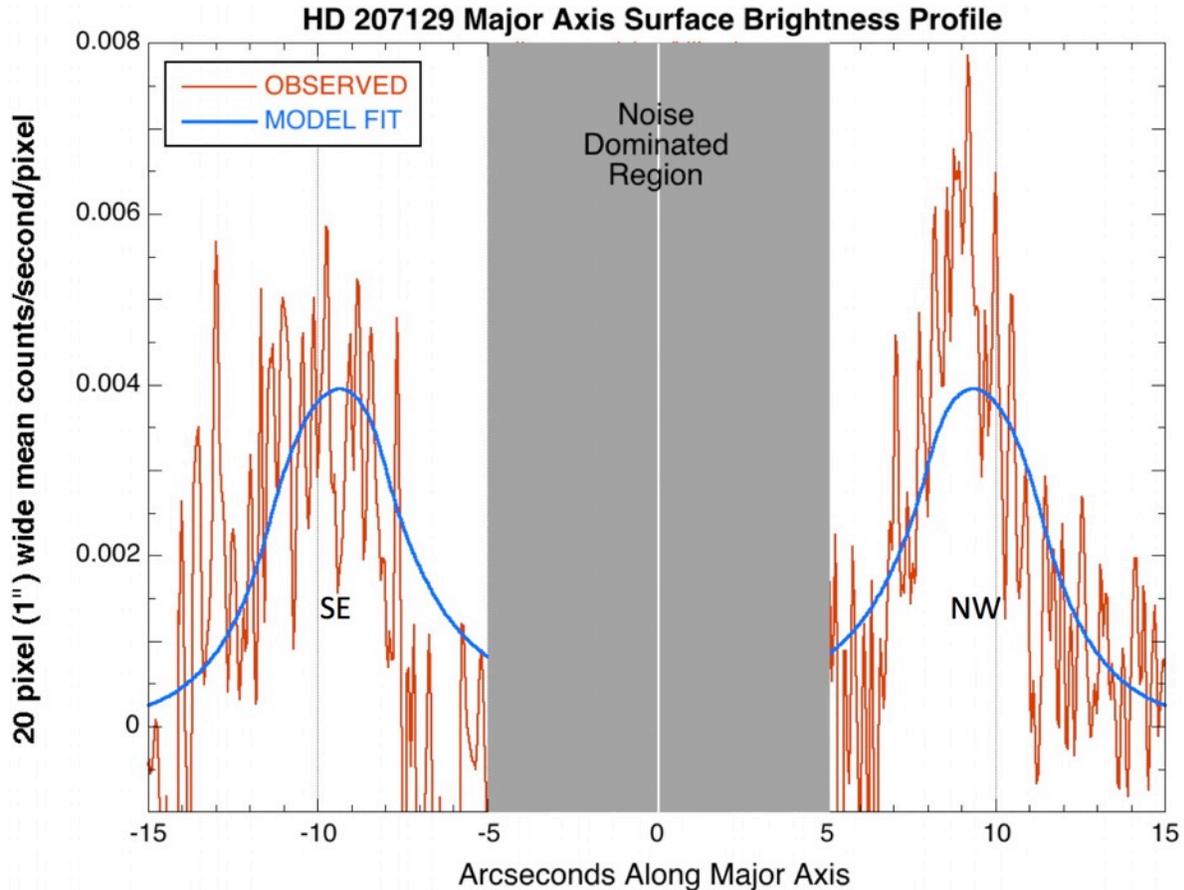

Fig 3. Radial SB profile of the HD 207129 CDS along the disk major axis (red), and as modeled (blue) as in Fig. 2 panel B, with peak SB ~ 24 $V_{mag}$ arcsec$^{-2}$ at ~9.5". At r ≤ 5" residuals from stellar light, incompletely suppressed with imperfect PSF-subtraction, dominates the stellocentric background increasing in brightness from ~ 23 $V_{mag}$ arcsec$^{-2}$ at 5" inward as illustrated in Fig. 4.

2. Non-Isotropic ("Front/Back") Scattering Asymmetry: From an analysis of the 2006 epoch ACS imaging of the HD 207129 CDS, Krist et al. (2010) suggested "at 0.6 μm, the ring shows no significant brightness asymmetry, implying little or no forward scattering by its constituent dust". They asserted $g$ < 0.1 with an interpretive consequence that the "nearly isotropic scattering in the ring conflicts with any assumption of spherical particles that scatter according to Mie theory" (separate from a low albedo inconsistency for silicate grains). The STIS data, in contrast, inform directionally preferential anisotropic scattering that, with $g \approx 0.25$ from the best-fit modeling, is visually apparent as a "front/back" asymmetry (see Fig. 2 panel A). In validation, though noisier, this same "front/back" asymmetry was seen in the STIS data when the first and second epoch data were separately reduced into two independent three-roll images originating at different field orientation angles. The reason for this discrepancy in *g* with the prior published ACS imaging is unclear. We speculate this may be related to a lower SNR (with shorter total integration time, narrower bandwidth, and lesser coronagraphic throughput) in the ACS data, or with the differing methods of reduction. The ACS image was derived from "roll subtraction", using the host star as its own PSF template. This obviates chromatic effects in PSF-subtraction but, as noted prior, is subject to possible localized disk-flux loss from self-subtraction with dependency on stellocentric distance varying with azimuth angle; i.e., differently in the ring at the disk minor vs. major axes. The disparity may be more related to image structure (i.e., fidelity), rather than photometric efficacy, as both the STIS (6 roll) and ACS (2 roll) reduced data sets have relatively good agreement in the 0.6 μm total disk flux density and average ansal SB.

3. Radial Sensitivity to Starlight-Scattering Material:

(a) Interior (endo-ring) dust: We do not detect any light-scattering dust (either diffuse, or in a secondary ring) comparable in brightness to the debris ring itself to an inner limiting stellocentric angle of ~ 5". To quantitatively estimate the imaging sensitivity limits to any (undetected) CS dust in the presence of incompletely suppressed starlight at, and interior to, this region, in Fig. 4 we plot the 360° median absolute deviation[7] (MAD) in the color-corrected PSF subtraction residuals as a function of stellocentric angle after subtracting the Fig. 2 panel B debris ring model. Very similar results are obtained with direct measurement of the stellocentric field, masking the regions along the disk minor axis that are superimposed upon the cento-symmetric residual pattern.

---

[7] With a relatively small number of rolls (maximum 6) contributing to each AQ image pixel, the MAD is a more robust (against outlier) estimator than the oft-used standard deviation; MAD ≈ 1.48 σ for large samples.

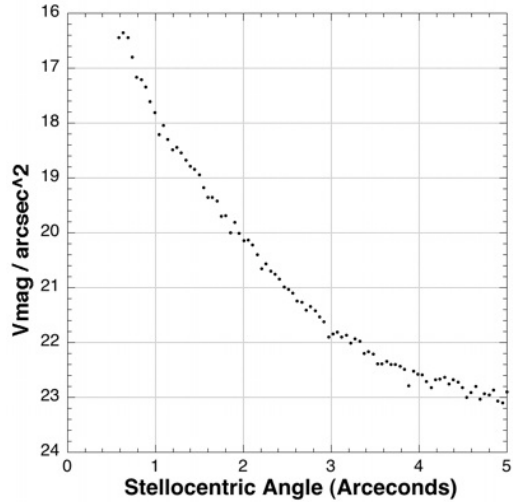

Fig 4. Median absolute deviation (MAD) limiting sensitivity for the detection of light-scattering debris dust as imaged in the inner stellocentric 5" of the HD 207129 CDS. Measured from the full set of color-corrected PSF-subtracted images at each roll angle observed.

(b) Exterior (exo-ring) dust: We also do not detect any light-scattering material beyond the exo-ansal radial SB decay as shown, e.g., along the disk major axis in Fig. 3. Beyond the ring seen in sky-plane projection, the background in the sectors toward and flanking its semi-minor axes at r ≥ 9–10" are fully unaffected by the presence of the interior disk and, then, exterior is radially invariant. We directly assess the imaging sensitivity limits in these regions (representative of fully circum-azimuthal measures) that are not dominated by PSF-subtraction residuals via multi-position aperture photometry as $\approx 25.8 \pm 0.3$ $V_{mag}$ arcsec$^{-2}$.

(c) Intra-ring dust: r ≈ 9 – 10" is also the distance to the ring peaks along the major axis, and so we approximately assess an intra-ring ansal average SNR ≈ 2.4 arcsec$^{-2}$.

## 7.2 – HD 202628

*Introductory Notes.* HD 202628 is one of ten nearby (< 40 pc) solar analog (K0 - G5) stars that were surveyed by Krist et al. (2012) for the presence of starlight-scattering CS dust potentially detectable with *HST* visible-light coronagraphy. This sample was IR selected with Spitzer-derived excess thermal emission $L_{ir}/L_{star} > 10^{-4}$, indicative of cold dust at stellocentric distances potentially within the imaging sensitivity capabilities of STIS for systems with favorable viewing geometries, dust properties, and ring-like morphologies with inner radii beyond the inner working angle limits of the coronagraph configuration employed. HD 202628 is a nearby (24.4 pc) solar-type (G2V) star with a thermal infrared excess $L_{ir}/L_{star} = 1.4 \times 10^{-4}$ (Koerner et al. 2010) estimated from 70 μm Spitzer/MIPS photometry, but lacking an excess at 24 μm from which Krist et al. (*ibid*) postulated an inner radius for the emissive dust of 80 au (4.1"). Of the ten stars surveyed, HD 202628 was the only one for which a visible-light debris system was imaged. The discovery imaging revealed a low-SB, intermediate-inclination, ring-like disk. Krist et al. (*ibid*) estimated a semi-major axis length to the debris ring inner radius of ~ 158 au, with characteristic width ~ 58 au averaged radially across both ansae, and noted a significant stellocentric offset.

*Observations and PSF Subtraction.* On 2015 May 30 and September 17, *HST*/STIS coronagraphic images of the HD 202628 CDS were obtained to improve, and expand upon, the discovery mode imaging of Krist et al. 2012. The discovery images were: (a) obtained using STIS occulting Wedge-A at its 1.8" wide location, (b) acquired in two *HST* orbits differing in a single orientation differential of 28°, (c) with total integration time 4512 s, (d) without contemporaneously observed, nearby, color-matched PSF template stars, (e) reduced with two-roll "self-subtraction" of the underlying stellar PSF. The GO 13786 follow-up images were: (a) obtained using STIS occulting wedge-A at a combination of its (narrower) 0.6" and 1.0" wide locations, (b) acquired in six *HST* orbits with five field orientation differentials spanning 262° (see Table 3), (c) with total integration time 12670 s, (d) with target-specific PSF reference stars, (e) reduced with 6-roll PSF template subtraction.

All observations executed as planned at both observational epochs. However, the pre-planed PSF template star for the first observational epoch (visit 07), BX Mic, was found "polluted" by third-light from an *a priori* unrecognized close-proximity companion (or ill-placed background star) rendering it unsuitable as a PSF-subtraction template. For the second epoch observations, the PSF template star was changed to an alternate candidate, HR 8042, and found to be an excellent match. Fortuitously, PSF-subtraction using the second epoch HD 8042 imaging showed little (if any) degradation due to possible differential breathing when applied to the first epoch (visit 05, 06, 08) observations of HD 202628, and so was used as a PSF template for both sets of HD 202628 visits. We applied the same method of image reduction, and two-wedge six-roll visit combinations of PSF-subtracted images as discussed for HD 207129 (§ 7.1), to produce the analysis quality (AQ) data image (shown in Fig. 5) discussed herein.

*Principal Results.* The STIS 6R-PSFTSC AQ image of the HD 202628 CDS (Fig. 5), from visual inspection, is generally in good morphological and photometric agreement with the Krist et al. (2012) discovery image (*c.f.*, their Fig. 2 top panel), though we note and discuss below some differences in details. The 6R-PSFTSC AQ image directly informs of the spatial distribution of the CS dust at all azimuth angles at r > 3.5" – 4.0", of greater apparent complexity than HD 207129. Principle geometric, photometric, and astrometric results discussed are summarized in Table 5.

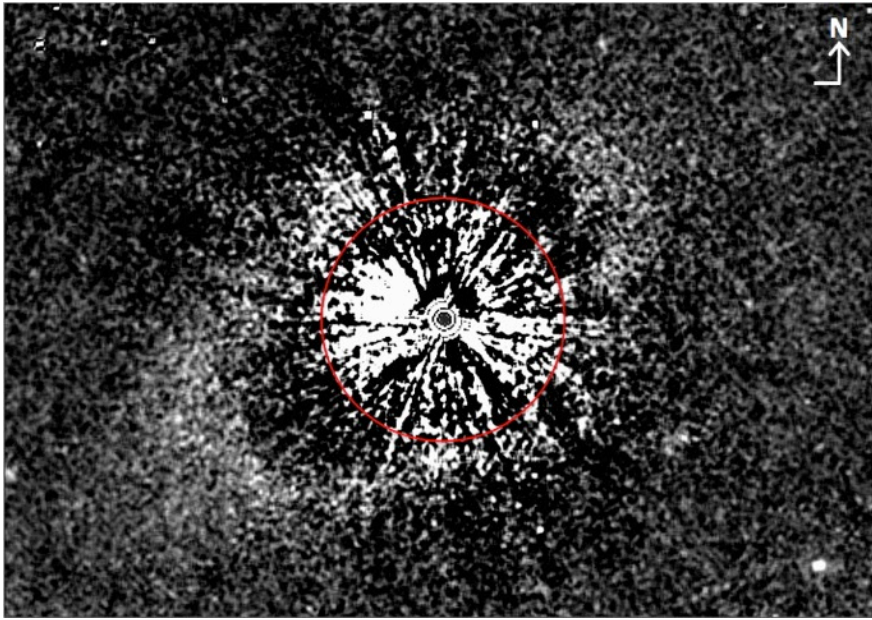

Fig 5. STIS 6R-PSFTSC image of the HD 202628 CDS. FOV 500 x 350 pixels (25.4" x 17.8" ) centered on the host star. Linear display stretch -0.002 (hard black) to +0.009 (hard white) counts $s^{-1}$ pixel$^{-1}$ with 3x3 boxcar smoothing to better illustrate the nature and amplitude of PSF subtraction residuals interior, and structure of the background exterior, to the debris ring. The red circle, with r = 3.5", indicates the region within where stellocentric PSF-subtraction residuals closer to the star in dominate over the brightness of the apparently elliptical debris ring observed at larger stellocentric angles.

1. Morphology: The HD 202628 CDS appears to be comprised, in morphological description, by two main components. First, dominating the system, a stellocentrically offset debris ring, and second, a low-SB diffuse "cloud" of exo-ring scattering material seen both exterior and somewhat interior only to the SE ansa.

(a) The dust-scattered starlight from the ring appears to be mostly confined (within 50%-peak SB inner and outer isophotes) in an elliptical annulus between ~ 135 au and 218 au from the host star as seen in sky-plane projection. Interior to the ring, there is a notable deficit of scattering material in the two diametrically opposed sectors flanking the major axis (P. A. = 124°) from 17° < P. A.° < 158° and 271° < P .A.° < 354° to an inner working angle limit of ~ 4" where residual starlight begins to dominate the background interior to the debris ring. The debris ring itself deviates from bi-lateral symmetry with its geometrical center offset from the location of its host star.

(b) A notable low-SB excess of diffuse scattering material is seen externally, and somewhat superimposed upon and internally, to the SE ansa (not apparent in the Krist et al. 2012 discovery imaging), but with none seen at or beyond the diametrically opposed NW ansa; see Fig. 8 panel B). This "cloud" of material has a comparatively very shallow outer slope leaving the outer "edge" of the ring beyond the SE ansa not as sharp nor well defined as for the NW ansa. It may also apparently "intrude" (in morphological superposition) a short distance interior to the brightest part of the ring itself. The eastern sector of the ring between the NE minor axis (at $P. A. = 34°$) CCW to near the SE ansa is marginally brighter (and/or wider at comparable brightness) than the diametrically opposed sector.

(c) Krist et al. 2012 found the SB of the ring's SE ansa itself about half as bright as the NW ansa. This is an anisotropy that is not reproduced in the follow-up imaging presented here (e.g., see the major axis radial profile in Fig. 7). We speculate that the prior reported ansal SB asymmetry and dearth of detectable exo-ring material beyond the SE ansa in the discovery imaging may have been a consequence of two orientation roll-subtraction partially cancelling the diffuse dust signal at these conjoined locations. This conjecture may eventually be tested with additional multi-roll imaging from *HST* GO program 13455 (Krist 2015; priv. communication).

(d) As is also discernable in the prior discovery imaging, the apparent geometrical center of the debris ring in the follow-up imaging is offset to the east (toward $P. A. \sim 60°$) from the location of its stellar host. However, in the follow-up imaging no (expected) corresponding pericentric "glow" is seen. This suggests the possibility of an azimuthally non-uniform dust density distribution (with or without non-isotropic scattering) that, elsewhere, may be also locally implicated by the presence of the SE primarily exo-ansal diffuse cloud of material.

2. Geometry: Beginning with a first visual estimation of the debris-ring geometry from the AQ image, we then parametrically fitted the ring-like component of the CDS to the same model described in § 6. Our best fits to the simple ring models, illustrated in Fig. 6, inform a less steep intermediate inclination of ~ 54° from face-on, as opposed to ~ 64° estimated prior by Krist et al. (2012) from apparent major:minor axial length ratios. These authors noted that in the discovery imaging most of the disk in the direction of its minor axis was obscured (remediated in the follow-up 6-roll imaging reported here). We conjecture that with this lack of additional information there may have been a bias in the original inclination estimation. We also note a small (~ 6°) difference with the prior estimation of the celestial position angle of the disk major axis, derived as a free parameter in visual ellipse fitting with an *a priori* adopted inclination that had been fitted only to the inner edge of the elliptical annulus in the discovery imaging. While no formal errors were given by Krist et. al 2012 for these, and other estimated parameters, they note that reliable least-squares fitting was precluded due to low SNR. Given the above we (somewhat subjectively) do not find these differences discrepant with significance.

3. Stellocentric Offset: The originating set of GO 13786 images at each of the six field orientation angles, before PSF subtraction, record the *HST* diffraction spikes centered on the occulted star. Following Sch14, using the "X marks the spot" diffraction-spike fitting method, we are readily able to determine the location of the occulted star in the 6R-PSFTSC AQ image with an uncertainty of ~ ± 4 mas rms. We directly compare this to the geometrical center of the best-fit ring models to ascertain the presence and amount of a stellocentric offset of the debris ring that may inform a forced eccentricity (e.g., by an unseen planet). Krist et al. 2012 identified such an offset in the discovery imaging estimated from ellipse fits to the debris ring inner edges (only). We confirm such an offset from the follow-up imaging wherein we find from our best-fit ring model the apparent geometrical center of the debris ring offset from the star by $(\Delta X, \Delta Y) = [-0.47", +0.28"]$ (measured in a Cartesian coordinate system defined by the debris ring major and minor axes seen in sky-plane projection); I. e., an apparent offset of 0.545" toward $P. A. = 239.4°$. With sky-plane deprojection, assuming an inclination of 54.2° from face on, this is a deprojected offset of 0.67" (16.4 au). This offset, determined in fitting the entire debris ring ellipse, while closely agreeing in celestial $P. A.$, is smaller than estimated by Krist et al. 2012 using the inner edge only in the discovery imaging.

4. (Nearly) Isotropic Scattering: With an assumption of an azimuthally uniform dust density distribution, a pericentric brightening of the debris ring of ≈ +13% is expected toward $P. A.$ ≈ 240° (the direction from the debris ring center toward the host star), and is encoded in the isotropic ($g = 0.0$) scattering model illustrated in Fig. 6 (middle panels). This, however, is not seen in the observed image (top panel), and indeed it is the diametrically opposed sector toward P. A. ≈ 60° that is marginally brighter. Such a brightening, independent of the $r^{-2}$ scaling expected with offset stellocentric distances, suggests the possibility of a dust density enhancement to the NE (though we cannot rule out some contribution to the possibility of some directionally preferential scattering). This is also consistent with contiguously adjacent brightening by the low SB "cloud" of material seen beyond the eastern ansa and may (or may not) be a physically contiguous feature. The ring itself is relatively well-fit, over the explorable range of scattering phase angles, assuming nearly isotropic Henyey-Greenstein scattering, $g = 0.1 \pm 0.1$, with the forward scattering direction along the ring minor axis to the NE with residuals shown for both flanking cases in Fig. 6 (bottom panels). After subtracting off the $g = 0.0$ model, the residuals along the ring are enhanced relative to $g = 0.2$ to the north of the star (toward scattering phase angle zero) and diminished to the south, and vice-versa. (This is readily apparent when "blinking" between these two images). Those residuals, however, are (globally) not fully eliminated with intermediate asymmetry parameter values also modeled. In any case, the residuals to the SE of the star, near and beyond and beyond the SE ring ansa are (not surprisingly) dominated by excess light from the previously discussed exo-ring "cloud" that is not included in the simple ring model.

5. "Clumpiness" Along the Ring? A careful examination of the spatial distribution of the residuals in the location of the ring, after subtracting the best-fit model, reveals some "clumpy" substructures in particular along its southern side (see Fig. 6 bottom panel) on multi-pixel spatial scales that one would not expect from an unperturbed (smooth) dust density distribution. These clumps may simply be imaging artifacts at the faint-limit of image fidelity - though comparable posited artifacts are not seen elsewhere in the field, or possibly confusion from ill-placed background sources. In principle, with 4 years of stellar proper motion of ~ 1.0", the Krist et. al 2012 discovery image could be used to arbitrate the latter. However, the same sub-structure pattern is not apparent in discovery-epoch, roll-subtraction, image (*c.f.*, their Fig. 2 top panel). This is likely due (if "real") to a combination of poor roll angle coverage (wedge obscuration) in the sector(s) flanking the disk minor axis, and differences in imaging efficacy with roll-subtraction vs. multi-roll

PSF template subtraction. At this time a possible intrinsic origin for the intra-ring substructures seen in the follow-up images cannot be confirmed, but is testable with comparable (or improved) third-epoch imaging in the future.

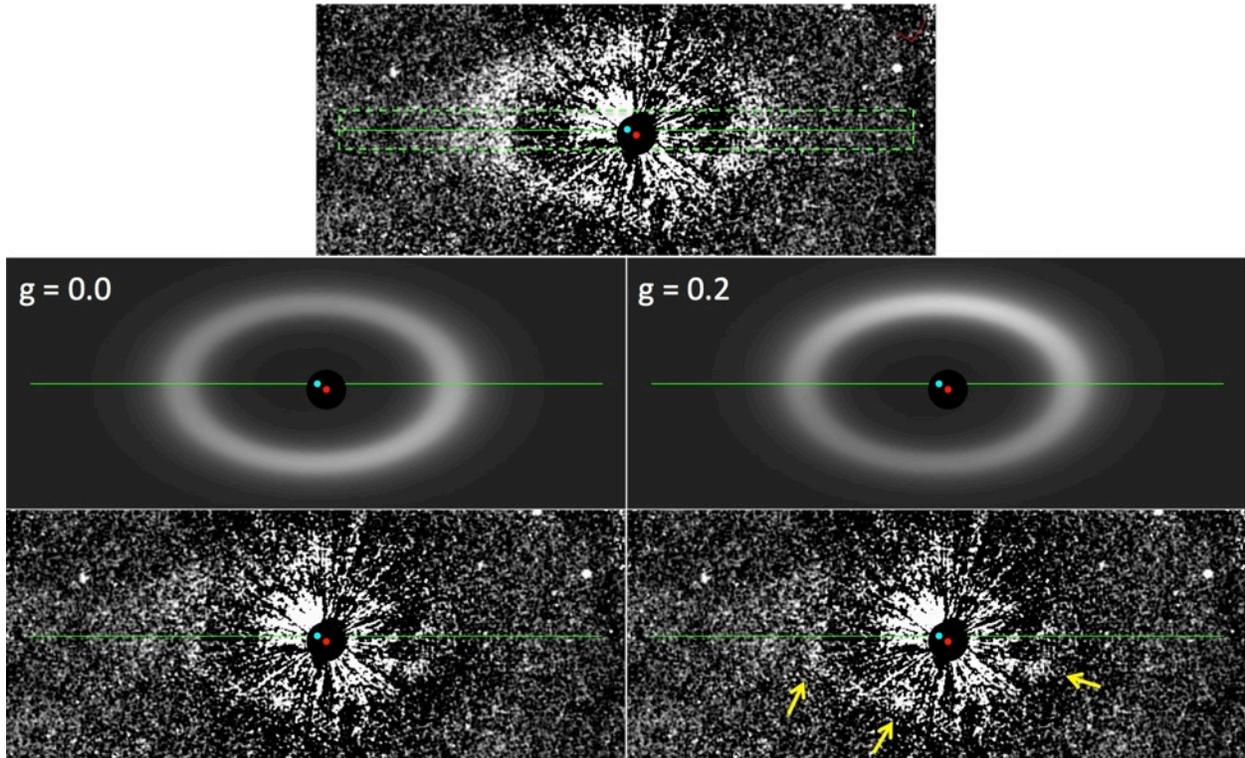

Fig 6. Observed (top), ring model (middle), and observed-minus-model residual (bottom) images of the HD 202628 debris system. All images are linear gray scale from -0.001 to + 0.005 instrumental counts s$^{-1}$ pixel$^{-1}$, north up, with FOV 33" x 14.5". The green dots mark the fitted location of the debris ring center, the red dots mark the measured location of the occulted star. The horizontal green lines mark the ring major axis. The dashed green rectangle on the observed image indicates the region used to measure the radial SB profile along the major axis as shown in Fig. 7. Yellow arrows indicate some of the "clumps" seen at the location of the debris ring after subtracting the ring model.

6. Major Axis Radial SB Profile: The major axis radial SB profile of the CDS, presented in Fig. 7 (black line), was measured along the disk major axis in both directions extending ± 15" (~ ± 300 pixels) from the center of the debris ring. The rectangular measurement region with its long dimension parallel to the debris ring major axis is depicted in Fig. 6.

(top panel). Along that strip we measured the average flux density in single-pixel wide bins, ± 20 pixels in extent orthogonal to the disk major axis, at stellocentric angles > 5" (beyond the central noise-dominated region). In Fig. 7 we also overplotted the identically measured major-axis radial SB profiles of the best-fit ring-models previously discussed with characterizing parameters given in Table 5. Along the major axis (and therein through the diametrically opposed ansae) the difference in the $g = 0.0$ and $g = 0.2$ models are indistinguishable at the resolution of Fig. 7. Here, we constrained the models to reproduce (scale) the SB of the ring at the NW ansa. The ~ 26% difference in the predicted diametrically opposing ansal flux densities arises from the stellocentric offset of the debris ring. The NW side of the ring along the major axis is well-fit to the model (with a noted deviation, however, at 7.5" < r < 8.7"). The model, in the presence of the exo-cloud clump on the SE side of the CDS, under-predicts the SB, though the inner edge location (defined by the 50% intensity point) and outer "edge" slope (though offset in brightness) are well matched to the observations.

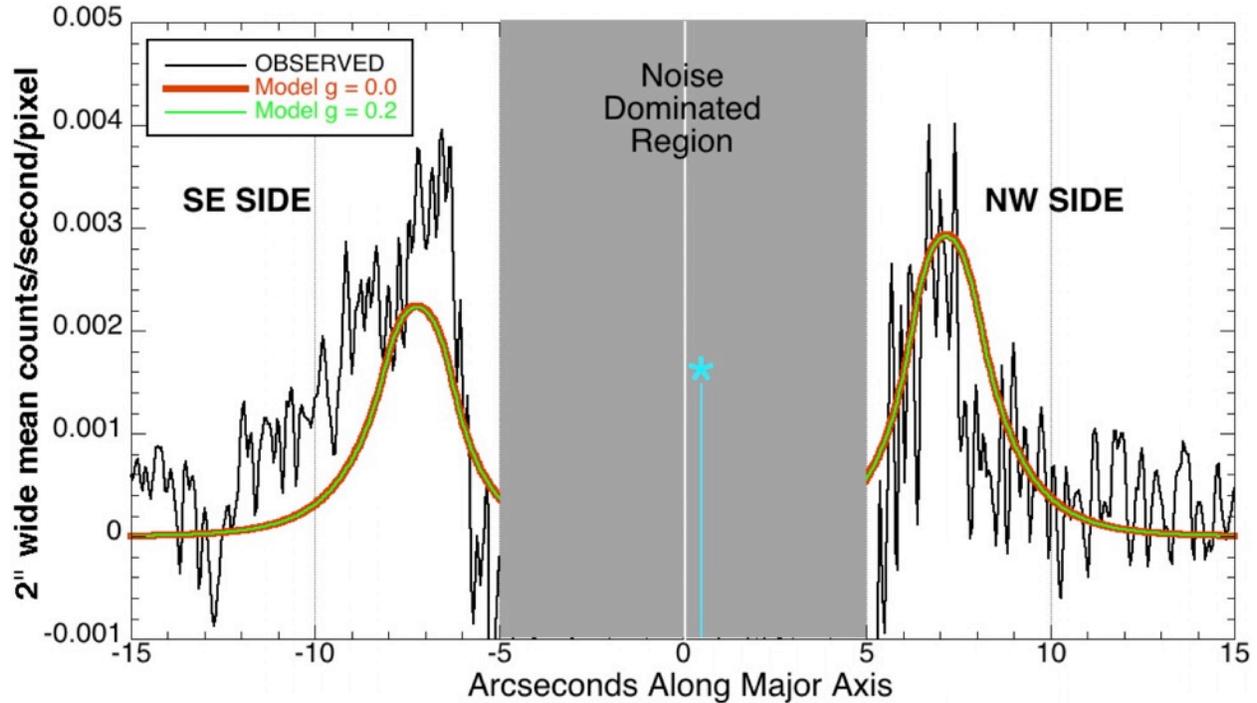

Fig 7. Radial SB profile of the HD 202628 debris ring with respect to the geometrical center of the of the ring ellipse along its apparent (sky-plane projected) major axis - with models constrained to fit the peak intensity on the NW side of the star (see text).

7. Ring Width and Edge Slopes: Following Schneider et al. 1999 (in the case of HR 4796A), we characterize and measure the inner and outer radii of the debris ring along its major axis from the locations where the radial SB has declined to 50% of its peak. We additionally estimate the SB gradient across the inner edge of the ring, i.e., its slope, fit with a radial power law dependence along the debris ring major axis. For these (steep) inner edge slopes, we fit radial regions from 142 to 158 au (~ 5.8" to 6.5") on both sides of the star centered on the disk. Radially beyond the radii of peak brightness, we separately fit in three contiguous regions between 175 au and 300 au (see Table 5). These metrics, in combination with the ring diameter, may be used to set constraints on the locations and masses of unseen planets that maybe responsible for the ring architecture; e.g., see Rodigas et al. (2014a). The presence of the low SB cloud of material producing additional scattered light at and beyond the SE ring ansa bias the adjoining outer slope and thus width measurements at and should be viewed with caution. Physical modeling, beyond the scope of this current paper, is necessary to disentangle the two.

**Table 5**
HD 202628 Debris Ring Parameters

| | | | | |
|---|---|---|---|---|
| Inclination (*i*) from face-on | 54.2° ± ~ 4° (axial ratio: 1.708:1) | | | |
| P. A. debris ring major axis | 124° ± ~ 3° (E of celestial N) | | | |
| *a* (elliptical annulus of peak SB) | 7.19" | | 175.3 au | |
| | | | | |
| 50% EDGES & PEAK RADII | FROM STAR | | FROM DISK CENTER | |
| NW inner-edge (major axis) | 6.409" | 156.4 au | 5.940" | 144.9 au |
| NW major axis peak SB radius | 7.628" | 186.1 au | 7.159" | 174.7 au |
| NW outer-edge (major axis) | 8.948" | 218.3 au | 8.479" | 206.9 au |
| | | | | |
| SE inner-edge (major axis) | 5.522" | 134.7 au | 5.991" | 146.2 au |
| SE major axis peak SB | 6.740" | 164.5 au | 7.209" | 179.5 au |
| SE outer-edge (major axis) radius | 8.060" | 196.7 au | 8.529" | 209.9 au |
| | | | | |
| SB SLOPES (Power Law Index) | SE (DISK CENTERED) | | NW (DISK CENTERED) | |
| 142 AU < R < 158 AU (inner) | +12.6 | | +7.7 | |
| 175 AU < R < 200 AU (outer) | -4.5 | | -12.9 | |
| 200 AU < R < 250 AU (outer) | -3.9 | | -2.8 | |
| 250 AU < R < 300 AU (outer) | -6.9 | | --- | |
| | | | | |
| NW Ring width (FWHM; Δr) | 2.539";  60.0 au | | | |
| g (H-G scattering asymmetry) | 0.1 ± 0.1 | | | |
| Stellocentric offset from ring center in sky-plane projection* | +0.469" (ΔX), –0.277" (ΔY) = 0.535" toward ring-azimuth 239.4° | | | |
| Stellocentric offset from ring center in face-on deprojection | 0.683"   16.7 au | | | |
| | | | | |
| SB @ NW ansa | 0.40 μJy arcsec$^{-2}$ | | 24.6 V$_{mag}$ arcsec$^{-2}$ | |
| SB @ SE ansa | 0.51 μJy arcsec$^{-2}$ | | 24.4 V$_{mag}$ arcsec$^{-2}$ | |
| Total CDS flux density (F$_{disk}$) | 44.6 ± ~ 9 μJy | | | |
| Disk (0.6 μm) Scattering Fraction[a] | 7.65 x 10$^{-6}$ | | | |

\* measured along the disk major (X) and minor (Y) axes

8. Exo-Ring Extent: Fig. 7 also shows that the major axis radial SB profiles across both ring ansae are comparatively asymmetric, in particular beyond the radius of peak brightness. The observations inform (different from Krist et al. 2012) similar ansal peak SBs on opposite sides of the disk. Except for a local deficit between ~ 7.4" and 8.5", on the NW side the model reproduces the outer-edge slope and brightness fall-off to the level of the sky noise at r < 11" (~ 270 au). On the SE side the disk, in the presence of a local exo-ring brightness enhancement visible in Fig. 5, the model under-predicts the brightness beyond the ring ansa to ~ 12.5" (~ 305 au) on (and flanking) the major axis. This may be evidence of exo-ring scattering material seen preferentially only beyond the NE ansa. Speculatively, such material may arise from the launching of small grains after an episodic collision, or from density perturbations ("confinement" by resonances) by unseen planet(s). Higher (still) SNR observations to better constrain the SPF would be beneficial to disambiguate the possibilities.

9. Total Disk Flux Density & Scattering Fraction. Following chromatic correction in PSF-subtraction, a faint, but not non-insignificant, azimuthally- and cento-symmetric negative residual, diminishing with stellocentric distance, remains present beyond the ~ 3.5" radius where highly-structured PSF-subtraction residuals otherwise dominate the background (see Fig. 8, panel A; dark area beyond the digital mask). The innermost r < 3.5" region (shown in Fig. 5) is digitally masked in making direct photometric measurements of the CDS. However, the presence of this remaining radial bias in the stellocentric background exterior, though of only minor consequence at the locations of the ring ansa at r ~ 7.2", by inclusion in the directions of and flanking the ring semi-minor axes of length r ~ 4.2" correspondingly biases aperture photometry. We therefore use the best-fit parametric model of the ring to estimate its total brightness, exclusive of the SE exo-ansal extension that is not included in the ring model. We re-fit with the same geometrical parameters of the ring, but to separately match the brightness of the ring in only in the diametrically opposed 90° sectors flanking the major axis where the radial background gradient is a much smaller bias. In this manner we roughly estimate that the full-ring fit under-estimates the total ring flux by ~ 30% in total flux density. With this systematic correction we then estimate the total instrumental brightness of the debris ring as 88 counts s$^{-1}$ ± ~ 20% or 4.0 ± 0.8 x 10$^{-5}$ Jy. This does not include the additional light scattered by the SE exo-ring "cloud". To estimate that contribution to the total brightness we subtract the best-fit ring model from the observed image and directly measure the (somewhat subjectively bounded as shown in Fig. 8) region contributing an additional ~ 4.6 ± 0.9 x 10$^{-6}$ µJy.; i.e., $F_{disk}$ = 45 ± ~ 9 µJy. We adopt a catalog $V_{mag}$ for HD202628 of 6.75 Jy (5.83 Jy in the STIS 50CCD band), from which the optical scattering fraction of the disk, $F_{disk}/F_{star}$ ≈ 7.7 x 10$^{-6}$, for spectrally flat gains.

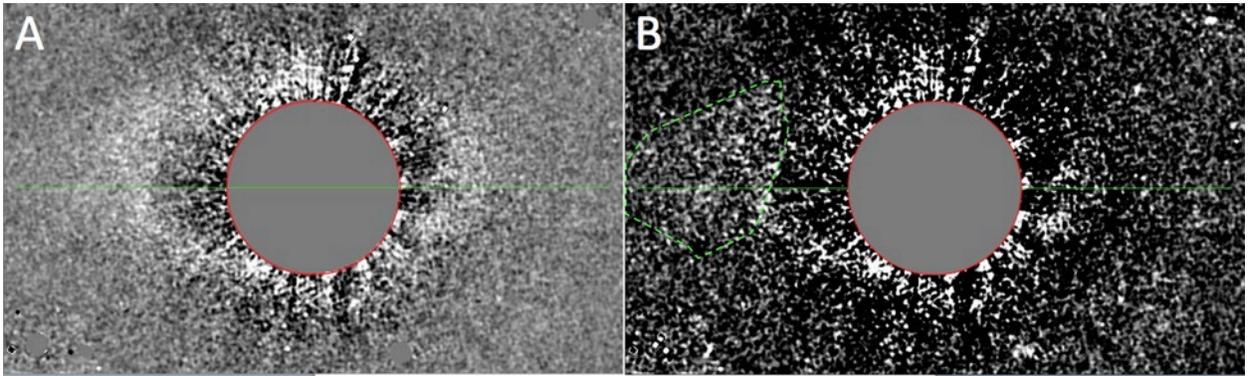

Fig 8. A: Enclosing region used to estimate the debris ring flux density in the presence of a residual chromatic bias (over-subtraction) in the stellocentric background with the r ≤ 70 pixel region dominated by PSF-subtraction residuals (show in Fig. 5) digitally masked. 500 x 300 pixels (25.3" x 15.2"; north up.) Linear gray scale display range ± 0.009 count s$^{-1}$ pixel$^{-1}$ to best show the background bias. B: After subtracting the best-fit model (stretch: -0.001 to + 0.005 count s$^{-1}$ pixel$^{-1}$) the region used to assess the flux density in the exoansal SE cloud is dashed in green.

10. Debris Ring SB. We characterize the representative SB of the CDS, independent of asymmetries expected from a stellocentric offset of the debris ring and non-isotropic scattering, by measuring and averaging the SB at both ansa. We do so in circular regions, 1 arcsec$^2$ in area, centered on the radial SB peaks at the ring ansa. At this stellocentric distance along the disk major axis (~ 7.2") the residual chromatic error in the background corrected for in the estimation of the total disk flux is small to negligible. We find (averaging the two ansal brightnesses differing by ± 12% with the NW side brighter) a mean SB of ~ 0.46 μJy arcsec$^{-2}$ (or ~ 24.5 V$_{mag}$ arcsec$^{-2}$).

### 7.3 – HD 202917

*Introductory Notes.* HD 202917 (HIP 105388; G7V, 43.0 pc) is a member of the Tucana-Horologium stellar association with an estimated age of 10 – 40 Myr (Moór et al. 2006). The presence of a thermally emissive CDS associated with HD 202917 was first suggested by Silverstone 2000 from 60 - 100 μm ISOphot data, later confirmed from its ~ 3 x 10$^{-4}$ infrared excess determined from Spitzer/MIPS observations (Beichman et al. 2005). An asymmetric scattered-light excess to the north (only) of the star was suggested by Krist et al. 2007 in "wide" V-band observations obtained in 2005 with ACS coronagraphy (*HST* program 10695) sensitive to dust-scattered starlight at stellocentric angular distance > ~1.5"; see Fig. 9 panel A as we have independently reduced those raw data obtained from the MAST archive with template PSF subtraction. Earlier (1999 in *HST* GO program 7226), NICMOS 1.6 μm coronagraphic images of HD 202917, designed for point-source (giant planet companion) detection by two-roll "self" subtraction, were obtained without PSF template observations. Later (2005 in *HST* GO program 10849; one of 21 targets), images of HD 202917 at 1.1 μm were acquired for disk-detection, but without a contemporaneous or specifically programmed PSF template star. In neither case was a scattered-light excess from a CDS detected. More recently, Soummer et al. 2015 re-reduced those prior data with principal component analysis (PCA) using a now-available extensive suite of PSF reference data from the MAST-hosted[8] (NICMOS coronagraphic) Legacy Archive PSF

---

[8] https://archive.stsci.edu/prepds/laplace/ in the Mikulski Archive for Space Telescopes (MAST)

Library (Schneider et al. 2011). The Soummer et al. (*ibid*) PCA reduction at 1.1 µm (that provides the best NICMOS spatial resolution) is reproduced in Fig. 9 panel B, wherein the northern "half" of the disk (well-correlated in position with the ACS image) was detected. From this image morphology, Soummer et al. 2015 suggested the "disk exhibits an asymmetric arc suggestive of a partial ring inclined ∼ 70° to the line of sight with a major axis *P. A.* ≈ 300° {=120°}", also noting that, "the northwest side is significantly brighter and more extended than the southeast side." Our new STIS PSFTSC imaging (Fig. 9 panel C, and Fig. 10) confirms these findings while providing both a higher-fidelity view of the ring-like disk and de-conflates now well-seen exo-ring material from light scattered by the debris ring itself.

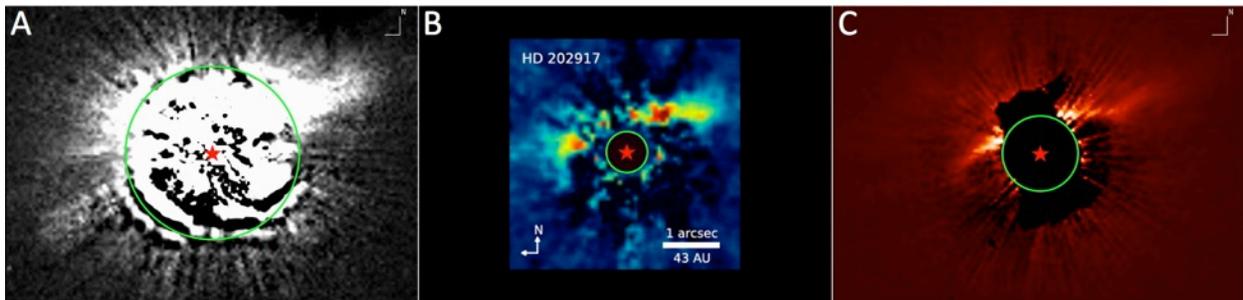

Fig 9. HD 202917 CDS. A) ACS F606W ("wide V" band) from re-processed *HST* program 10695 archival data. Scattered-light excess above the level of PSF-subtraction residuals is seen to the north of the star at r > 1.5". B) NICMOS F110W (1.1 µm) from Soummer et al. 2015. C) STIS unfiltered (optical broadband) with 6R-PSFTSC. Green circles indicate effective inner working angles: r = 1.5", 0.36", 0.66", respectively.

*Observations and PSF Subtraction (Details).*

WEDGEA-1.0: (*a*) All coronagraphic observations of HD 202917 from both the 2015 August 01(Visits 55 - 58) and October 13 (Visits 51 - 54) epochs executed nominally as planned with no absolute orientation constraints imposed. The HD 8893 WedgeA-1.0 PSF template observations obtained contemporaneously (as Visit 57) with the 01 August CDS images, however, were too deeply exposed. As a result, those images were saturated to larger-than-planned stellocentric angles in the directions orthogonal to the mid-line of Wedge-A to r ~ 1.1"

(22 pixels), and so do not, with application, provide information on the CDS at distances closer than this. However, the serendipitous absolute orientations for the relative roll constraints in these visits (but not the 13 October visits) had the Wedge-A itself obscuring the disk itself (so would not have been visible in these data in any case). (*b*) Exposure times for the second epoch HD 8893 PSF template visit (#53) were correctly adjusted prior to execution to circumvent saturation at r > 0.5" (10 pixels). With that, in the second epoch (only) PSF subtracted images, the ring-like disk is seen protruding on opposite sides of the Wedge-A at all three image orientations with a revealed major axis ring radius of ~ 1.5" (29.5 pixels).

WEDGEA-0.6: (*c*) The debris ring itself is also seen, but with inferior fidelity, in the less deeply exposed WedgeA-0.6 PSF-subtracted imaging for the more favorable (second epoch) visit 51-54 wedge orientations. With their PSF subtraction residuals, they contribute only marginally to improving the IWA over WedgeA-1.0 imaging from these visits. (*d*) Separately, the first-epoch (unfavorable wedge orientations) visit 57 WedgeA-0.6 PSF template star, while properly exposed, was (unfortunately) de-centered in acquisition w.r.t. contemporaneous target imaging. With PSF-subtraction this causes wedge-flanking instrumental brightness excesses (or deficits) obscuring stellocentric visibilities interior to the WedgeA-1.0 limits achieved, and contribute only "noise" in these smallest stellocentric regions.

*Analysis Quality Data Image.* For reasons (*a – d*) above, the Analysis Quality data images for Wedge-A1.0 and A0.6 for HD 207129 combines the PSF-subtracted data as follows:

(1) for WedgeA-1.0: (disk-favored orientation) visits 51 - 54 are used to an IWA limited by extent of the wedge,

(2) for WedgeA-1.0: (disk-disfavored orientation) visits 55 - 58 are used to a larger IWA limited by the PSF saturation distance,

(3) for WedgeA-0.6: (disk-favored orientation) visits 51 - 54 are used interior to regions unseen with WedgeA-1.0,

(4) for WedgeA-0.6: (disk-disfavored orientation) visits 55 - 58 are not used due to their inferior image quality.

N.B.: With respect to (4), the WedgeA-0.6 AQ images independently confirm the ring structure and morphology seen in the WedgeA-1.0. However, they additionally contribute little, and with noise actually do not improve, knowledge of the very small additional area unsampled with the 6-roll WedgeA-1.0 images. Therefore, in this case for quantitative measurements, only the WedgeA-1.0 6-roll combined AQ image is used.

*Roll Coverage*. The HD 202917 CDS is well-revealed in the STIS GO/13786 fully-reduced (6 roll combined) AQ image, as shown in Fig. 10 with the morphological major axis of its ring-like disk on the image horizontal. Due to *HST* schedulability constraints, the six field orientations satisfying other imaging needs to produce a high-fidelity AQ image could not also provide complete roll/IWA coverage needed to otherwise fully reveal the debris-ring at all CS azimuth angles close to the star. As a result, the debris ring at, and in its opposing sectors closely flanking, its apparent minor axis was not imaged (digitally masked as shown in Fig. 10). The unobscured majority remainder of the CDS, however, was imaged with high fidelity.

*Principal Results.*

1. Morphology: Unlike prior ACS and NICMOS re-processed images (Fig. 10, panels A & B), the STIS image clearly reveals a complex "two component" CDS morphology with: (1) a relatively narrow "half" ring-like disk that is apparently (at least partially) centrally cleared, and (2) a larger and fainter exo-ring "fan-like" starlight scattering structure. The latter is not unlike those seen in some other more highly inclined CDSs, e.g., HD 61005 and HD 32297 (Sch14) that, like the ring itself, is seen only one side of the star. Both "components" exhibit significant azimuthal asymmetries.

2. Non-Isotropic ("Front/Back") Scattering Asymmetry: The debris ring itself clearly exhibits a very strong "front/back" scattering phase-angle asymmetry, as suggested from the prior ACS and NICMOS imaging. The SB along the ring is as asymmetrically "mirrored" about the debris-ring major axis. This implicates strongly directionally preferential scattering, with (as a result) the fainter (southern) "half" of the debris ring undetected. The ring itself is seen to "turn around" to the south of both ansa, but only marginally so at its western extremity. The debris ring is undetected from stellocentric celestial $P.A. \approx 120° - 280°$. The exo-ring "fan" of material is, locally, significantly brighter beyond the ring's western ansa than on the diametrically opposed eastern side (see Fig. 10 panel B). The broad outer scattering "fan" seen only to the north of the star is morphologically skewed, by about 7° clockwise, with respect to the apparent celestial orientation of the debris ring axes (see Fig. 10 panel C).

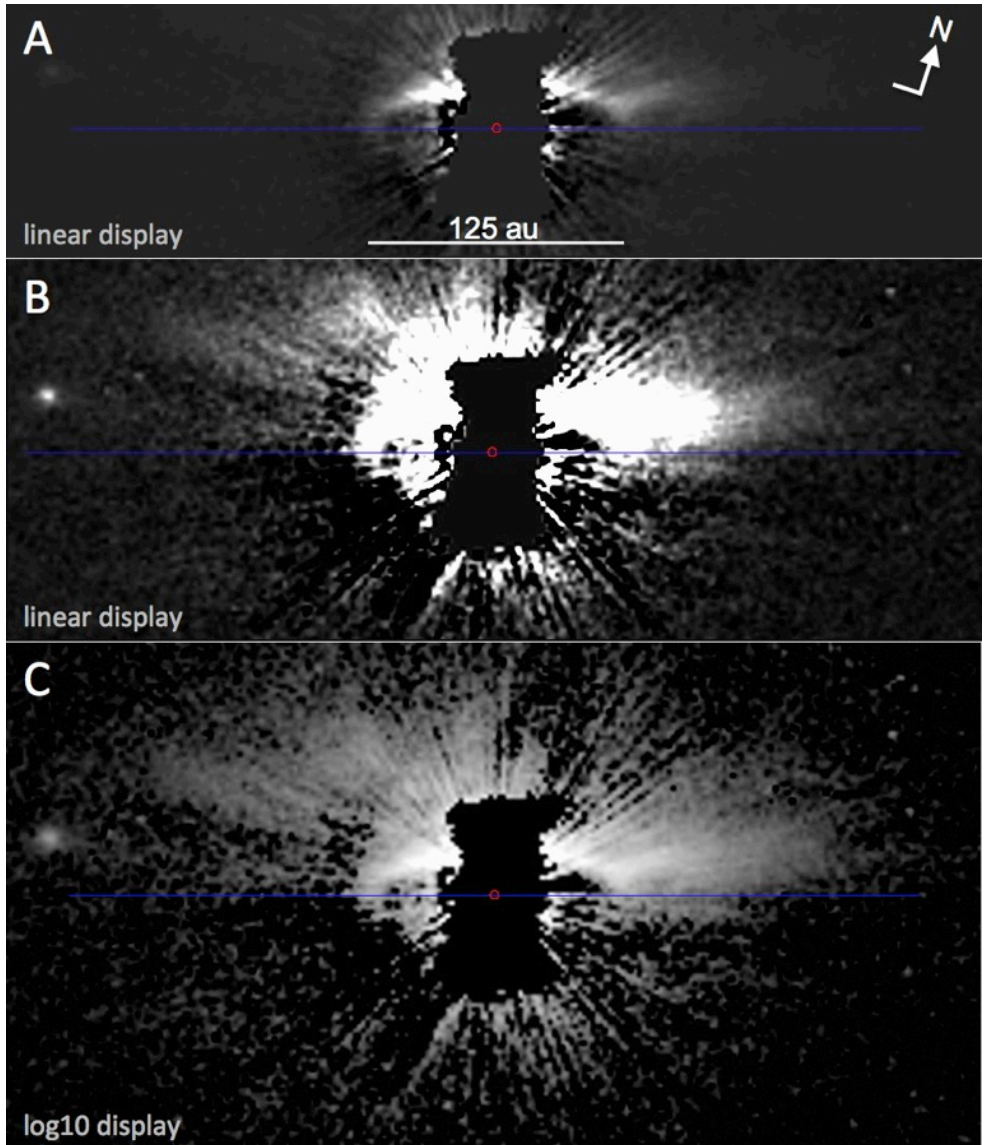

Fig 10. The highly asymmetric morphology of the HD 202917 debris ring and its exo-ring "skirt", both preferentially scattering (and seen) only one side of the ring major axis, is illustrated over a large a range of ~ 300x in SB in three display images (same data). Panel A from -0.01 to +0.1 count s$^{-1}$ pixel$^{-1}$ (FOV: 11.67" x 3.05"), panel B from -0.2 to +1.0 count s$^{-1}$ pixel$^{-1}$ (FOV: 11.67" x 4.57"), panel C from [-2.5] to [0] dex count s$^{-1}$ pixel$^{-1}$ (FOV: 11.67" x 6.09").

3. Model Fitting: The best fits to the ring-only model, minimizing the residuals along the visible portion of the debris ring after model subtraction, were found with a Henyey-Greenstein asymmetry parameter $g = 0.6 \pm 0.1$ with a correlated uncertainty in scaling the brightness of the model to the observed data by ± ~ 20%. In Fig. 11 we show these best fits with the model scaled in brightness by +20% with $g = +0.5$ and -20% with $g = +0.7$ (with the forward scattering direction toward the northern, visible, major axis) in Fig. 11. To be clear, we do not attempt to model the exo-ring "fan", but rather to disambiguate the two "components" of the CDS. Parametric results are summarized in Table 6.

**Table 6**
HD 202917 Debris Ring Parameters

| | | |
|---|---|---|
| Inclination ($i$) from face-on | 68.6° ± 1.5° from face-on | |
| P. A. Major Axis | 108° ± 1° E of N | |
| Ring peak semi-major axis ($a$) | 1.46" ± 0.02" | 62.8 ± 2.8 au [a] |
| Ring HWHM major axis inner-edge radius | 1.29" ± 0.02" | 55.4 ± 2.5 au [a] |
| Ring HWHM major axis outer-edge radius | 1.60" ± 0.02" | 68.6 ± 3.0 au [a] |
| Ring width (FWHM; $\Delta r$) | 0.13" ± 0.03" | 13.16 ± 1.3 au [a] |
| $g$ (HG scattering asymmetry parameter) | 0.6 ± 0.1 | |
| Debris ring flux density | 182 ± 36 c/s/pix | 83 ± 17 μJy |
| Total CDS (fan+ring) flux density [b] | ~ 660 c/s/ pix | 318 μJy |
| Total CDS (0.6 μm) Scattering Fraction [b] | ~ 3 x 10$^{-4}$ | |

[a] includes uncertainty in stellar parallax from SIMBAD
[b] Inner unsampled region obscured by coronagraph wedge estimated with best-fit model

4. Debris Ring Geometry: The observed ring geometry has only minor dependencies upon $g$ within its estimated uncertainty (see below). In Table 6 we give the best-fit parameters for $g = 0.6$ with systematic uncertainties included for solutions with $g$ varied by ± 0.1. Fitting the (visible portion) of the debris ring to the best-fit scattering models (Fig. 11) we find a semi-minor:major axial ratio ≈ 0.182 with a corresponding inclination ≈ 68.6° from a face-on viewing geometry for an (assumed) intrinsically circularly symmetric debris ring, with a celestial PA of its major axis of ≈108° east of north. We find the semi-major axis of the debris ring seen in sky-plane projection ≈ 1.46" (= 62.8 au) in length as the distance between the SB peaks along the major axis in the best-fit model. The characterizing width of the ring, based upon its inner-edge to outer-edge full-width to half intensity maximum is 13.2 au.

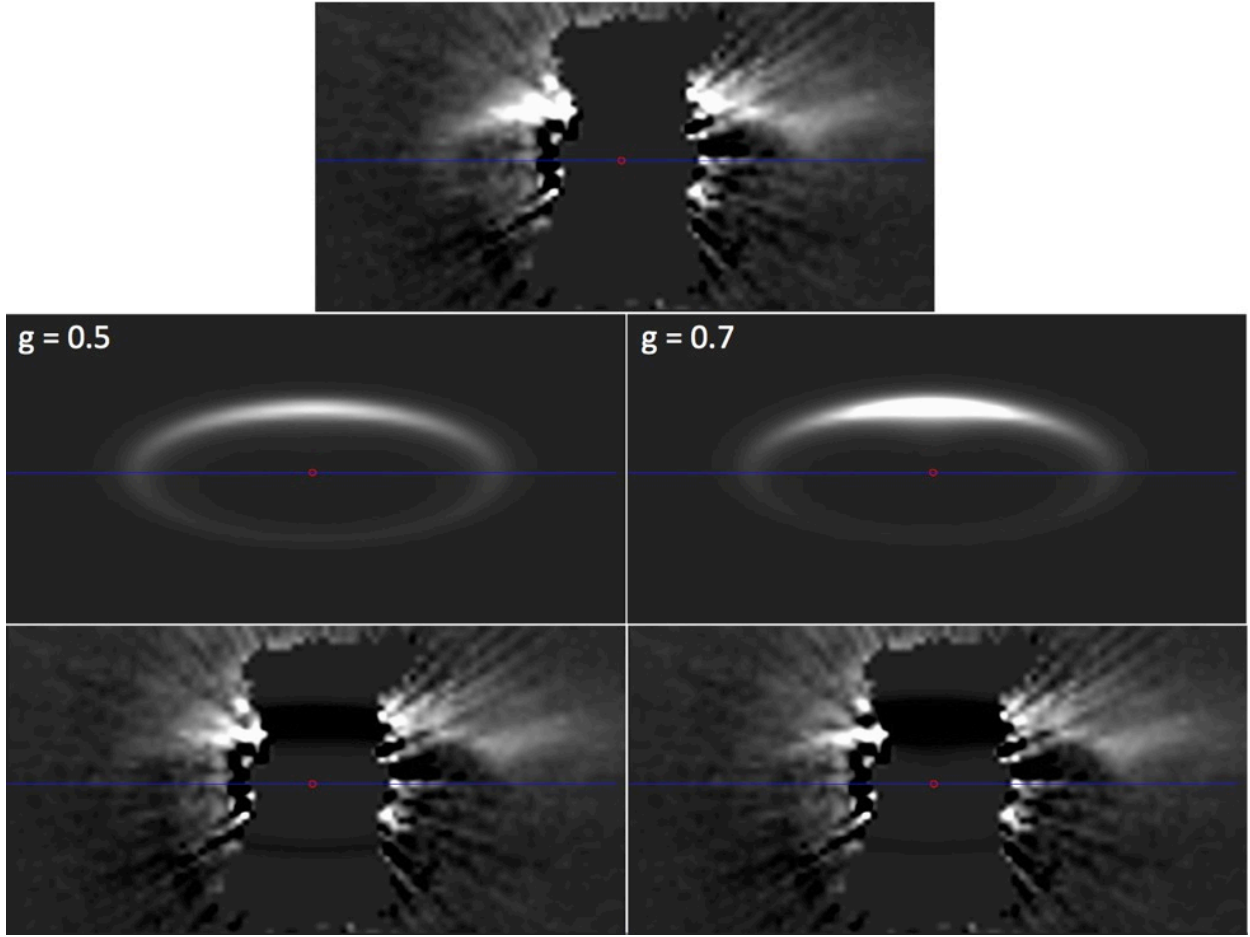

Fig 11. Observed (top), ring models (middle), and observed-minus-model residual (bottom) images of the HD 202917 debris system where significant anisotropic residuals remain (see text). All images are linear gray scale from -0.2 to + 1.0 instrumental counts s$^{-1}$ pixel$^{-1}$, north up, with FOV 5.07" x 2.53". Red dots mark the measured location of the occulted star. The horizontal blue lines mark the ring major axis.

5. Asymmetries and Anisotropic Scattering: We note two major scattering anisotropies in the HD 207129 CDS. *First*, the debris ring itself exhibits highly directionally preferential scattering (g ~ 0.6) with its brighter side toward its projected minor axis (Fig. 10, panel A). The STIS data, unfortunately, did not image the brightest part of the ring. However, the imaged regions along the northern ring-arc appear asymmetrically bright on opposite sides of the minor axis (eastern side brighter). This is perhaps better seen after subtracting off the bi-laterally symmetric ring models (best-fit in the 0.5 < g < 0.7 range), as shown in Fig.

10, panel C, where a bright feature superimposed at the location of the ring-arc remains on the eastern side of the unsampled area, but not so on the western side. This suggests there may be (some) azimuthally differentiated sub-structure along the debris ring. *Second*, scattering by exo-ring material on the western side of the CDS flanking (to the north) of the ring major axis is significantly brighter than on its otherwise mirror-symmetric eastern side (Fig. 10, panel B). This suggests a local enhancement in the surface density of scattering particles (through a recent collision or other causality) beyond the bright ring toward celestial PA ~ 300°. This was suggested previously by Soummer et al. 2015 from the PCA reduced NICMOS image, though the direction of the debris ring major axis that we now see is 12° further clockwise.

6. CDS Extent: Lower surface-brightness exo-ring material in the CS "fan" is seen over the same range of scattering phase angles as for is ring itself (suggesting, perhaps, some common physical material properties). This material is detected to a maximum stellocentric distance (+3$\sigma$ SB above the sky background) of ~ 5" toward celestial PAs ≈ 90° and 295°. Just to the north of the western debris ring ansa where the aforementioned exo-ring bright scattering region is locally seen. This is coincident with the region detected in ACS PSF-subtracted imaging from archival data processed from the MAST. This feature is also likely the origin of the suggestion by Soummer et al. 2015 that the northwest side is significantly brighter (than the SE side), though the significance in the ACS image is not so apparent.

7. Photometry: To estimate the total flux density of the HD 207129 debris ring only, superimposed on its embedding larger CDS structure, we use the best-fit scattering models of the ring with uncertainties in brightness scaling for the range of $0.5 < g < 0.7$. To estimate the total flux density of the full CDS (both the "skirt" and the embedded ring together) we perform area-aperture photometry fully enclosing the CDS and then use the superimposed portion of the ring-model coincident with the interior region of the CDS that was unsampled in the STIS image. Both are given in Table 6 showing that about 2/3 of the total CDS brightness originates in scattering by exo-ring material. With HD 202917 at $V_{mag}$ = 8.67, the visible light scattering fraction of the CDS, $F_{disk}/F_{star}$ is ~ 3 x $10^{-4}$. This is ~ 40x "brighter" (in $F_{disk}/F_{star}$) than both of the ~ 2.3 Gyr G-star disks discussed in § 7.1 & 7.2, and is approximately equal to its IR excess (2.5 x $10^{-4}$).

## 8. DISCUSSION – COMPARATIVE G-STAR DISKOLOGY

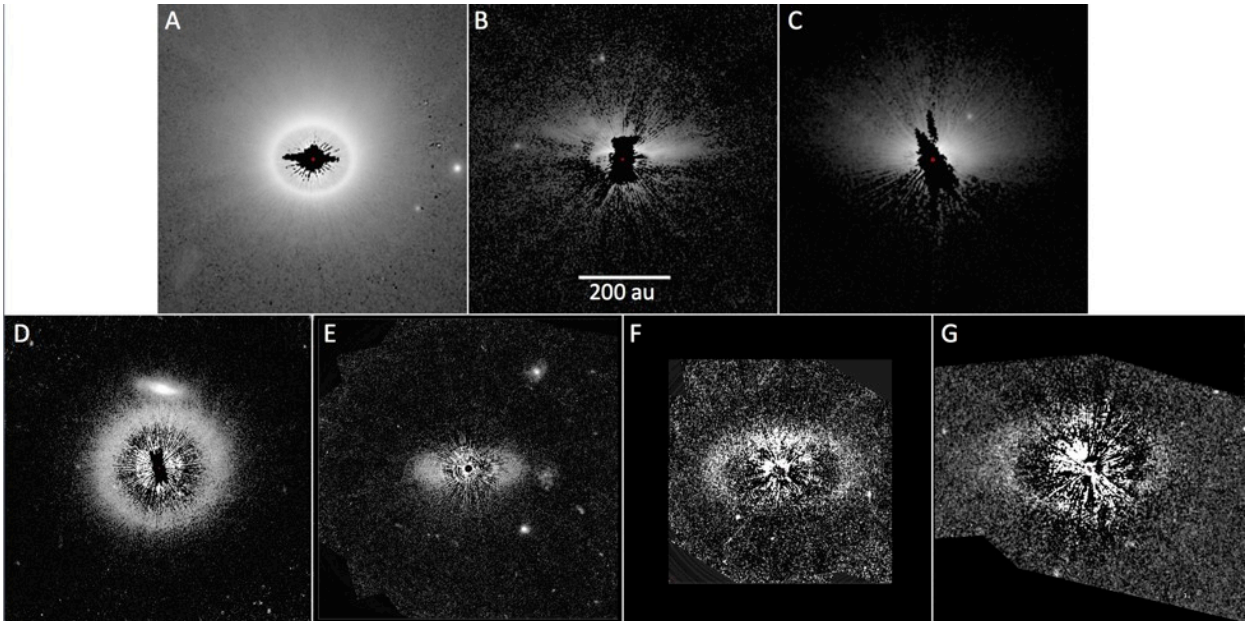

Fig 12. Intermediate-inclination CDSs with well-resolved ring-like components hosted by FGK solar-analog stars as imaged homogeneously with STIS/6R-PSFTSC are presented in increasing order (panels A-G) of stellar age (also in seen-correlated decreasing $F_{disk}/F_{star}$; See Fig. 13.) A: HD 181327 (F6V), B: HD 202917 (G7V), C: HD 15745 (F2V), D: HD 107146 (G2V, with bright background galaxy in the field), E: HD 92945 (K1V), F: HD 207129 (G2V), G: HD 202628 (G5V). A, C, D, and E from *HST* program 12228 (Sch14), B, F, and G from *HST* program 13786 as discussed in this paper. All are shown at the same physical scale in au (see scale bar applicable to all) with the major axis of their debris rings as seen in sky-plane projection oriented on the image horizontal.

In § 7 we presented in detail the reduced images and observationally derived metrics characterizing and describing the starlight-scattering debris systems circumscribing three G-type stars: HD 202917, HD 207129, and HD 202628, all with intermediate-inclination ring-like disks. All three systems are very well detected and resolved with STIS 6R-PSFTSC revealing debris rings that are several times larger than our solar system's Edgeworth-Kuiper belt whose inner edge is maintained by Neptune (Liou & Zook 1999). In the specific case of HD 202628 its stellocentric offset may imply a large planet many times more distant than Neptune is from the Sun.

In Fig. 12 we present images of these three G-star CDSs on a common physical scale, along with other spatially well-resolved solar-analog (FGK star) CDSs with intermediate inclination ring-like disks that we previously imaged in *HST* GO program 12228. Together, these cover an age range from ~ 23 Myr (panel A; HD 181327) to ~ 2.3 Gy (panels F & G). The latter, HD 207129 and HD 202628 are very mature CDSs of only half the solar age and more massive analogs of our own solar system's circum-solar Edgeworth-Kuiper belt that has persisted for ~ 4.6 Gyr.

### 8.1. $F_{disk}/F_{star}$ Brightness Over Time

HD 207129 (G2V) and HD 202628 (G5V), both ~ 2.3 Gyr in age, are the "faintest" solar-analog CDSs yet observed in terms of optical scattering fraction, $F_{disk}/F_{star}$, with both ~ 8 x $10^{-6}$. Despite their ansal SBs of only ~ 24.0 and 24.5 $V_{mag}$ arcsec$^{-2}$, respectively, they both are readily detectable and mapped with STIS 6R-PSFTSC in part due to their large major axis angular radii of ~ 9.3" and ~ 7.1". This is, in angular scale, beyond the region interior where PSF subtraction residuals dominate. The optical scattering fractions of these two oldest solar-analog CDSs are in stark contrast to the ~ 23 Myr HD 181327 (F6V) and ~ 90 Myr HD 61005 (G8V) with their ~

200 - 300 brighter $F_{disk}/F_{star}$.

The well-determined (≤ ~ 10% uncertainty) total disk optical scattering fractions of all seven intermediate inclination FGK-star ring-like CDSs that are shown in Fig. 12 are plotted in Fig. 13 (left panel) as a function of their estimated host star ages, along with that also of the nearly edge-on HD 61005 (Sch14 and Fig. 16). Age estimations and their uncertainties are taken from the literature; see Table 1 of this paper for HD 202917, HD 207129 and HD 202628, Mamajek & Bell (2014) for HD 181327, and Table 1 of Sch14 for the other stars[9] plotted in Fig. 13.

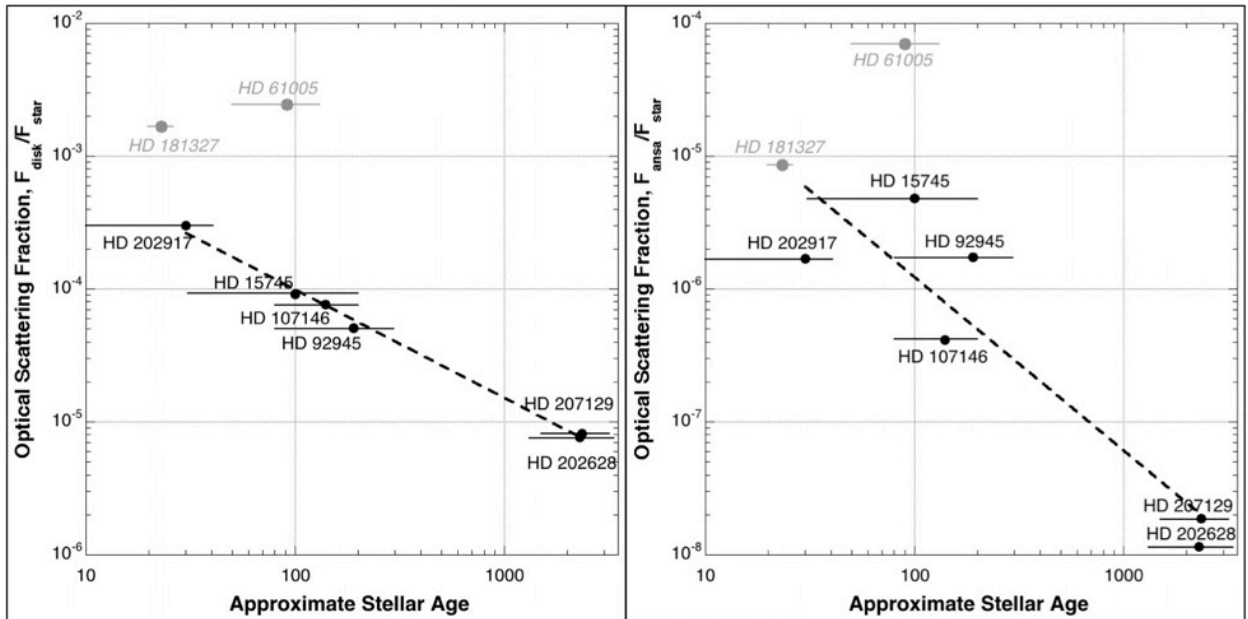

Fig 13. Left: Plot of $F_{disk}/F_{star}$ as a function of stellar age for the seven intermediate-inclination solar-analog hosted CDSs shown in Fig. 12, and the nearly edge-on G8V-host outlier HD 61005 (Fig. 16). All were measured from nearly identically-observed STIS STIS/6R-PSFTSC images. The dashed line is the formal best fit, $t^{-0.81}$, power law excluding outliers HD181327 and HD 61005 (gray points, see main text). Right: Similar spatially localized plot of the debris ring 100 au$^2$ ansal flux densities compared to the total stellar light and the best-fit (dashed line) power law.

Both HD 181327 and HD 61005 have been shown to possess very large, and optically-bright, exo-ring scattered-light halos well beyond the peak radii of their thermal IR excess emission (Sch14); For HD 181327 posited as arising from a recent massive collision in its birth ring (Stark et al., 2014), and for HD 61005 with small-particle mass-loss through a ram-pressure interaction with the local ISM (e.g., Hines et al 2006). Excluding these two systems with non-steady state dust production events, the remaining more quiescent CDSs are seen as an ensemble to decay in visible light scattering fractions, $F_{disk}/F_{star}$, over time (t) approximated by a $t^{-0.8}$ power law from ~ 23 Myr to 2.3 Gy. The robustness and more general applicability of this trend, however, should be viewed with due caution given the small sample size and origin. The fit is suggestive of an evolutionary optical dimming timescale for overall ensemble, but in general may be quite inhomogeneous for individual CDSs diverse in stellar host, disk, and environmental characteristics. Additionally, this follow-up sample was culled from sensitivity-limited discovery-mode survey imaging for CDSs with $L_{ir}/L_{star} \geq 10^{-4}$, and thus could be biased for systems intrinsically fainter in their excess IR emission. The efficacy of the suggested power-law timescale remains to be tested when a larger, and fainter, homogeneous sample of solar analog debris systems may be available in the future and is necessary to reliably establish a posited evolutionary sequence.

---

[9] For the uncertainty in the age of HD 15745 we adopt a lower value of ~ 30 Myr as suggested by Zuckerman & Song (2004), and for an upper-value a canonical age estimation as a likely/probable member of the Castor Moving Group of ~ 200 Myr (Barrado y Navascues 1998).

A similar, but somewhat less steep, decay in 24 μm IR excess emission, tracing larger particles, was predicted by Gaspar et al (2013) over the same time for a broader range of host-star spectral types with also less massive disks than our $L_{ir}/L_{star} \geq 10^{-4}$ *HST* down-selected sample. The modeled average disk mass in their population synthesis has not yet reached quasi-steady state and is therefore decaying at a slower rate. However, the more massive members of their population synthesis have, and show, IR decay slopes in agreement with the visible-light observations of the more massive HST-observed disks. Steeper slopes, more closely matching those seen in our visible-light solar-analog ensemble, are predicted by Gaspar et al (*ibid*; *c.f.* their

Fig. 6) for initially more massive disks.

For each of these spatially well-resolved systems, as an ensemble in Fig. 13 (right panel), we also compare the total CDS flux density to that of the flux density in a 100 au$^2$ (seen in projection) region of their debris rings at their ansal SB peaks also ratioed to their host-star's flux density ($F_{ansa}/F_{star}$). A somewhat steeper, ~ $t^{-1.3}$, ansal-only power law is seen. Though this is not as well correlated in age for the younger (< ~ 2 x 10$^8$ Myr) systems alone, a similar declining trend is seen with the inclusion of the (~ 10x) oldest systems. The larger dispersion about the $F_{ansa}/F_{star}$ trend line may be due to local anisotropies in the dust density distribution (clumping, deficits, or enhancements from dust production events) in the small (100 au$^2$) areas measured in any given system at any given epoch.

Both of the oldest close-solar analog (G-star) CDSs reported in this paper, HD 207129 and HD 202968, have similar IR excesses of ~ 1 x 10$^{-4}$ (see Table 1)[10]. This is only about an order of magnitude less IR bright then the ~ 100x younger stars in this sample (HD 181327 at 2.5 x 10$^{-3}$ at ~ 23 Myr and HD 202917 at 1.2 x 10$^{-3}$ at ~ 30 Myr) despite their advanced ages. Although no color information is available within the broad STIS spectral passband to assess wavelength dependence (see § 8.3), with $F_{disk}/F_{star}$ for both an order of magnitude fainter than their IR excesses suggests grains of low optical albedo in these two oldest starlight-scattering CDSs.

### 8.2 Morphology and Structure

Starlight scattered by exoplanetary debris at visible wavelengths traces the small (micron sized) particles in CS environments. When ring-like structures exist, their locations, widths, and inner/outer edge slopes may predictively be able to constrain the dearth/existence, locations, masses, semi-major axes and eccentricities of co-orbiting planets (Rodigas et al. 2014a). These small particles are also easily reactive to non-planetary perturbative forces intrinsic to the systems such as stellar corpuscular winds, radiation pressure "pushing" material outward, and Poynting-Robertson drag causing the decay of particles inward. CDSs should not, however, be considered as closed or isolated systems. There now exists strong and growing evidence for ISM interactions with the small grain materials at large stellocentric distances in some CDSs manifested with the visual appearance of bow/termination shocks as well as large "skirts" of exo-ring material, for example in HD 61005 (Hines et al. 2007) and HD 32297 (Sch14). The latter may also be the case for HD 202917 and HD 15745 (Fig. 14).

---

[10] This is three orders of magnitude brighter in $L_{ir}/L_{star}$ than our own solar system's circumsolar debris disk.

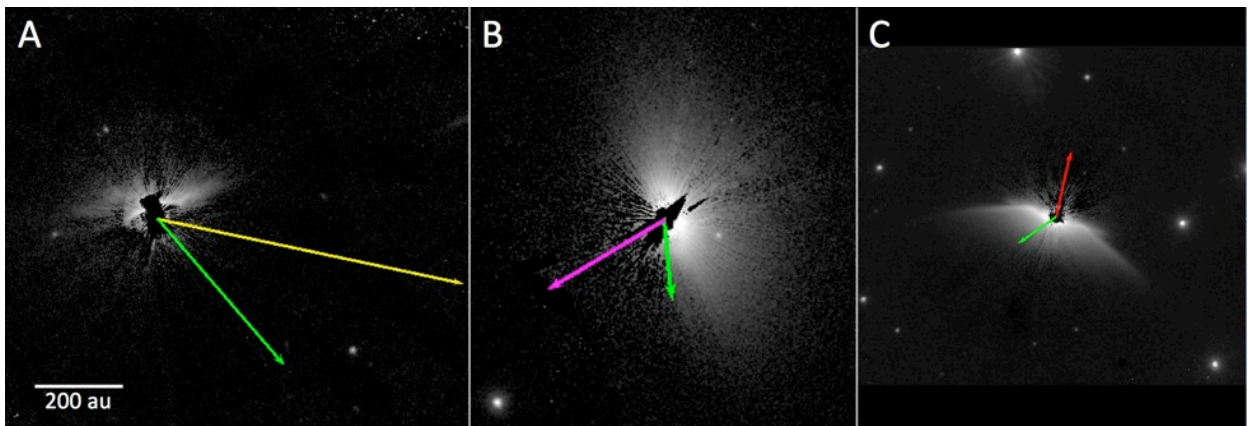

Fig 14. FGK-stars hosting CDSs with large skirts or fans of exo-ring materials seen only to one side of the star posited as "blown back" by ISM winds: (A) HD 202917, (B) HD 15745, and (C) HD 61005. Arrows illustrate the proper (tangential) motions of the CDSs with respect to those of the Local Interstellar Cloud (green), and with the other most likely interacting interstellar clouds at their stellar locations. The length of the arrows indicates the stellar/CDS motion through the ISM in 100 years. The most likely other interacting clouds identified from Redfield & Linsky 2008 are: For HD 202917, the Vel cloud with interaction velocity 7.1 au/yr (yellow arrow), for HD 15745 the Hyades cloud at 3.1 au/yr (purple arrow), and for HD 61005, the *"blue"* cloud (red arrow) at 1.5 au/yr. All images are presented north up, east left, at the same physical scale in au along their respective ring major axes.

### 8.2.1 Comparing HD 202917 to HD 15745 and HD 61005

We compare the morphology of the HD 202917 (Fig. 10) CDS with those of HD 15745 (Fig. 15) and HD 61005 (Fig. 16). All are in the likely middle-age range of ~ 30 - 90 Myr, with HD

202917 the youngest of the three. All are hosted by FGK stars, with HD 61005 (G8V) a very similar close solar-analog to HD 202917 (G7V). All three bear some significant morphological resemblances with large "skirts" or "fans" of exo-ring scattering material.

In Sch14, the HD 15745 CDS was discussed and presented to best illustrate its extended "fan"-like outer scattering structure (as so originally described from its ACS discovery imaging by Kalas et al. 2007), without emphasis on the debris ring itself. In Fig. 15 we show the HD 15745 CDS to better illustrate its intermediate inclination ($i \approx 57°$), r = 1.17" (74 au), debris ring within. The ring is a lower contrast feature, relative to its exo-ring "fan" (or "skirt"), than HD 202917 and HD 61005, and is obscured in a linear representation (Fig. 14 panel A, similar to Fig. 13 panel C of Sch14). With a logarithmic compression of the dynamic display range, panel B, the debris ring is readily apparent. In panel C we scale the SB in the plane of the ring to compensate for the $1/r^2$ dimming of stellocentric illumination, which further enhances the visibility of the ring and of the endo-ring region of the CDS relatively devoid of starlight-scattering material. While the region of the ring along its minor axis (in particular flanking its fainter semi-minor axis) was unimaged due to *HST* observing constraints, it is clear that the CDS exhibits highly directionally preferential scattering. This is also the case (more extremely) with HD 202917 at a somewhat similar mid (68°) inclination angle.

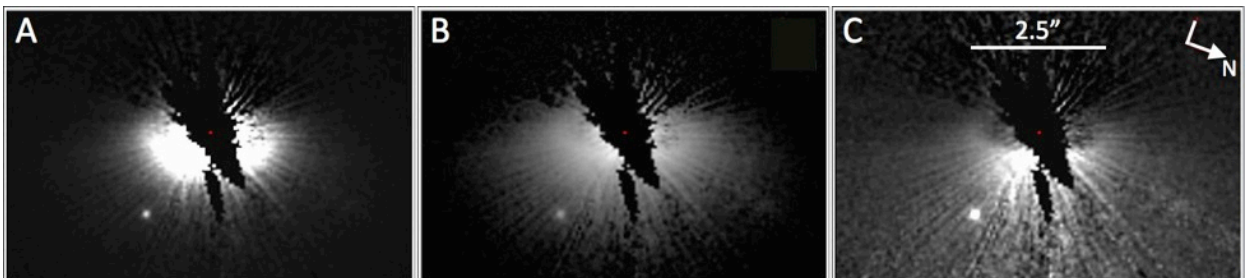

Fig 15. Revealing the HD 15745 starlight-scattering debris ring. Data from *HST* program 12228 as discussed in Sch14. A: Linear display from -0.1 to + 3.0 counts s$^{-1}$ pixel$^{-1}$. B: Log$_{10}$ display from [-1.0] to [+1.5] dex counts s$^{-1}$ pixel$^{-1}$. C: Linear display identical to panel A after scaling the SB to compensate for the $1/r^2$ falloff in stellar

illumination in the plane of the ring. Field of view: 7.6" x 5.1" all three panels.

While the HD 61005 CDS has a highly inclined (~85°) debris ring (unlike those in the Fig. 12 intermediate-inclination FGK star sample) it, like HD 202917's, none-the-less has been well-detected and resolved in scattered-light -- along at least along its brighter "half" (e.g., Buenzli et al. 2010; Sch14; Olofsson et al. 2016). Like HD 202917, a highly asymmetric scattering phase function is suggested as responsible for the non-detection of the fainter "half" of HD 61005's debris ring. Quite differently, however, at its likely intermediate age of 90 ± 40 Myr, the HD 61005 CDS sits well "above" the $F_{disk}/F_{star} \sim t^{-0.8}$ power law fit to the intermediate-inclination systems plotted in Fig. 13.

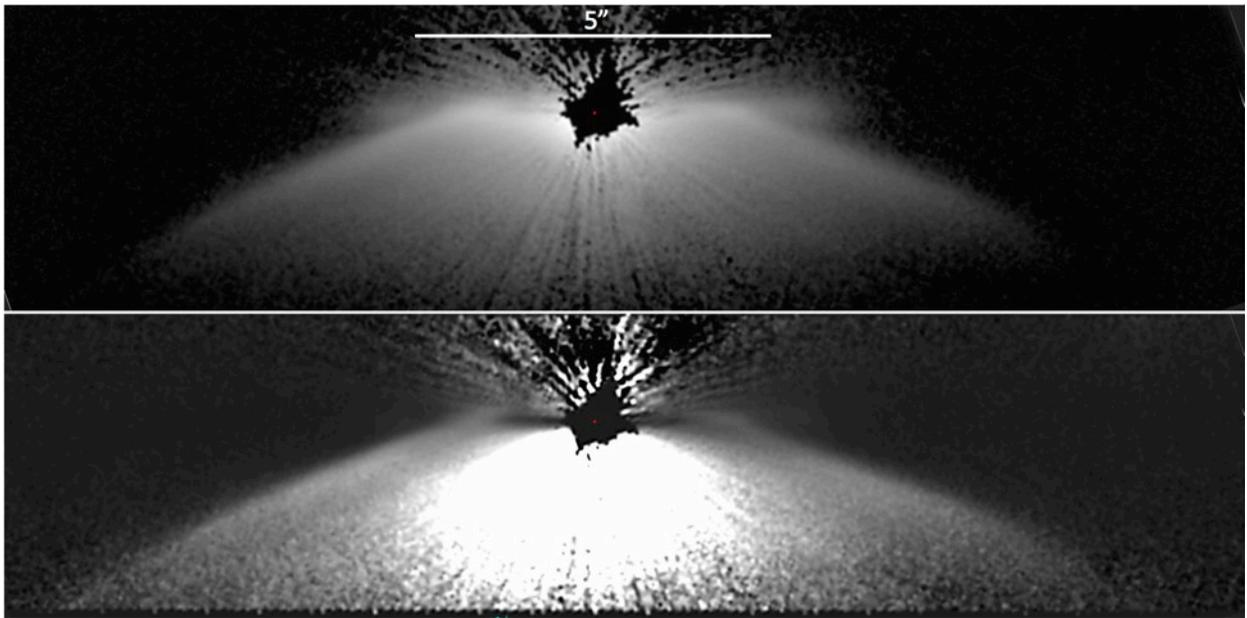

Fig 16. Top: HD 61005 CDS SB image, as discussed by Sch14, in a $\log_{10}$ display stretch from [-2] to [+1.5] dex counts s$^{-1}$ pixel$^{-1}$. Its ~ 54 AU radius debris ring is seen in sky-plane projection along with its exo-ring "skirt" of starlight-scattering material and its leading edge bow shocks on either side of the ring minor axis. Bottom: Compensating for the $1/r^2$ diminution of stellar light in the (highly inclined) plane of the ring, the central clearing of material in the endo-ring region is readily apparent in this linear display stretch from [-1] to [+10] {dex} counts s$^{-1}$ pixel$^{-1}$. This radial re-scaling of the brightness may over-enhance the azimuthal brightness asymmetry (non-isotropic

with scattering phase angle) within skirt of materials that may be non-coplanar with the debris ring (see footnote 12).

HD 61005's near edge-on inclination (like also HD32297's with a more massive A6V[11] host) contributes to the *HST*-detectability of its very large and asymmetric outer scattering structure (see Fig. 16, panel a) of small (micron size) starlight-scattering grains. Such asymmetric structures have been posited as arising from an ISM wind blowing back the small grains in the disk halo (Artymowicz & Clampin 1997). At our viewing geometry this is manifested as a large exo-ring "skirt" of scattering material seen in sky-plane projection. Its endo-ring inner clearing is very apparent when compensating in the ring-plane for the $1/r^2$ attenuation of the stellar illumination[12]; see Fig. 16 panel B. The stirring/launching/ejection of the starlight-scattering exo-ring grains by ISM interaction may contribute to this CDS's notable morphology and to its $F_{disk}/F_{star}$ brightness for its age, rather than, e.g., invoking a recent catastrophic collision.

The HD 202917 CDS, though less highly inclined to our line-of-sight than the HD 61005 CDS, bears morphological resemblance (at lower SNR) in its SB image (compare Fig. 10 panel C to Fig. 16 top). In both of these systems only the brighter "half" of the debris ring with a very large front-to-back scattering asymmetry is detected. Both, along with HD 15745, also possess large skirts or fans of exo-ring scattering material seen on the brighter side only of the CDSs. HD 202917 differs, however, from both HD 61005 and HD 15745 in that it is significantly asymmetrically brighter on one (the east) side of the ring minor axis at and beyond the ring itself.

---

[11] Recent suggested reclassification by Rodigas et al. 2014b with $T_{eff}$ = 8000±150K and M = 1.65±0.1$M_{sun}$.

[12] In this $1/r^2$ presentation, the in-plane radial illumination correction that collapses the imaging dynamic also enhances the visibility of azimuthally non-isotropic scattering around the exo-ring skirt. This is also the case for the HD 15745 exo-ring fan. However, as the exo-ring material may be non-coplanar with the ring, and indeed may possess a more complex three-dimensional scattering surface (e.g., an "umbrella geometry") seen in two-dimensional projection. Thus, a simplifying assumption of co-planarity may overly-enhance this anisotropy with scattering phase angle. While the exo-ring inclination geometry may not be correct as a function of stellocentric distance, the causality for the enhancement is probably a real effect. The scattering phase angle directly "below" the star on the system minor axis (r = 0, z increasing downward in cylindrical coordinates) should be the minimum scattering angle as a function of z, precisely where a HG scattering phase function is peaking. Hence, for a "fixed" z (ignoring inclination) a peak in SB directly below the star is expected. Detailed modeling beyond the scope of this paper (Hines et al. 2016, in prep) can constrain the inclinations of the exo-ring light-scattering material.

The morphology of such a structure may be difficult to reproduce through an ISM interaction alone, and may be indicative of a recent dust enhancement event through a catastrophic collision or other dust production event. Kalas et al. 2007 disfavored ISM blowback as causal for the HD 202917 fan, noting the star's tangential motion and location well within the local bubble. However, a plausibly analogous bow shock seen in IR observations of δ Vel (Gaspar et al. 2008), also in the local bubble with similar relative motion to the ISM, may suggest otherwise and is indicative of the fact that higher density cloud patches exist in the local bubble. Reproducing the morphology of the HD 202917 CDS is a challenge to models of disk structural evolution inclusive of both forces intrinsic and extrinsic to the system (beyond the scope of this current paper).

### 8.2.2 Inter-Comparing the Oldsters – HD 207129 and HD 202628

We also inter-compare the two, approximately equally oldest (~ 2.3 Gyr), G-star CDSs yet imaged, HD207129 (G2V) and HD 202628 (G5V). Both share similar ring-like morphologies, are roughly comparable in size ($a$ = 149 au, and 175 au, respectively), and have very similar optical scattering fractions (~ 8 x $10^{-6}$). The HD 207917 CDS exhibits a high degree of bi-axial morphological symmetry, with its azimuthally varying SB well fit with a Henyey-Greenstein scattering model with g ~ +0.25. HD 202628 appears axially mirror symmetric about its apparent major axis (only) with only a small scattering phase asymmetry ($g$ ~ +0.1). The HD 207129 debris ring appears centered on its stellar host, while HD 202628 is stellocentrically offset. No exo-ring material was detected exterior to the soft outer edge of the HD207129 debris ring, while a diffuse "cloud" of low SB exo-ring scattering material appears extending outward from the stellocentrically more distant ansa only of the HD202628 debris ring. This (so far) unique CDS asymmetry may (or may not) be associated with the stellocentric offset of the debris ring itself.

The combination of its stellocentric offset, and its diffuse exo-ansal cloud give HD 202628 its distinctively unique "teardrop" morphology. Additionally, after subtracting a SB model of the ring-like component of the HD 202628 CDS, a number of intra-ring "clumps" are revealed. While some (or even all) of these may arise from background contamination, their spatial distribution preferentially superimposed upon the location of the debris ring itself, and absent elsewhere, suggests that at least some of these are likely real from local dust density enhancements and not image artifacts. Reproducing these features in the HD 202628 CDS that do not appear with HD 207129 may present a challenge to future models of these two systems with very similar stellar hosts.

### 8.3 Fundamental Photometric Data for Modeling Material Properties

Both filter-band color diagnostics and directionally preferential scattering derived from spatially resolved imaging photometry can inform or constrain the physical properties of the starlight-scattering materials (e.g., Debes et al. 2008a,b). The former, however, cannot be obtained with STIS coronagraphic observations alone. However, the flux density and SB measures in the STIS 50CCD spectral band that we report in Tables 4 - 6 may be combined with future data to be obtained with continuing improvements to ground-based high-contrast imaging systems in the near-IR, and with the advent of JWST/nirCAM coronagraphy. Separately, photometrically determined scattering efficiencies as a function of stellocentric illumination phase angle with spatially resolved imaging, even in a single spectral band, parameterized by $g$, can provide a useful constraints in ascertaining the physical properties of materials in some CDSs. Unlike face-on or edge-on CDSs, the intermediate inclination CDSs we focus on in this paper simultaneously provide an explorable range of circum-azimuthal scattering phase angles (lacking in face-on disks) that (with sufficient resolution and image contrast) are spatially resolvable with significance and without self-obscuration of debris structures (in edge-on disks).

In fitting the STIS 50CCD band SB distributions of the ring-like components of the HD 202917, HD 207129 and HD 202628 intermediate inclination CDSs we assumed a simple, single characterizing HG asymmetry parameter, $g$; see § 6. More completely descriptive scattering models, e.g., those considering optical properties dependent upon a diversity in material composition, sizes, and mixes, are beyond the intended scope of this paper. Also see, e. g., Stark et al. 2015 for a discussion of the necessity for alternative or more complex formulations for some CDSs as explored in part for HD 181327. For the three G-star CDS rings we have observed, over the range of explorable scattering phase angles[13], we found a range from nearly isotropic scattering (HD 202628, $g \approx 0.1$) to mildly asymmetric scattering (HD 207129, $g \approx 0.25$) in the case of the two oldest CDSs, to strongly asymmetric scattering (HD 202916, $g \approx 0.6$) such that only the brighter side of the debris ring (centered on $\varphi = 0°$) was detectable.

## 9. Summary and Conclusions

(1) We have used STIS 6-roll PSF-template subtracted coronagraphy to explore the morphological, photometric, and astrometric properties of the endo-, intra-, and exo-ring environments of three G-star hosted circumstellar debris systems. STIS 6R-PSFTSC provides unique access to large (> a few arcsecond), and spatially diffuse, CDS sub-structures at low SB and high contrast beyond the outer working limits of current state-of-the-art high-contrast AO-augmented imaging systems (e.g., GPI high-contrast half-angle field limit = 1.7"). Imaging data were obtained in total light at visible wavelengths and without reliance on flux non-conservative reduction methods (such as ADI commonly used on the ground for contrast-augmentation) that otherwise often "self-subtracts" low-contrast diffuse structures increasing uncertainties in

---

[13] HD 202628: $\varphi = \pm (35.8°$ to $144.2°)$, HD 207129: $\varphi = \pm (32.1°$ to $147.9°)$, HD 202917: $\varphi = \pm (21.4°$ to $158.6°)$

morphological and photometric interpretation.

(2) For HD 207129, one of the two comparably old ($\approx$ 2.3 Gy) stars in our sample, we find a bi-axially symmetric ring-like debris disk exhibiting HG-like scattering with $g = 0.25 \pm 0.05$ (differing from Krist et al. 2010 discovery imaging). The debris ring peaks in radial SB at a stellocentric distance of $\approx$ 150 au, with a characteristic width of $\approx$ 73 au, and may be marginally brighter at its NW ansa compared to its SE ansa. Interior to the ring, with an inner (FWHM) edge radius of $\approx$ 116 au, we find no additional evidence of starlight scattering dust to limiting sensitivity in surface brightness of ~ 23 $V_{mag}$ arcsec$^{-2}$ at r = 5" (80 au) from the star, nor beyond its outer periphery to a limit of $\approx$ 25.8 $\pm$ 0.3 $V_{mag}$ arcsec$^{-2}$. The total optical light scattering fraction of the disk compared to the star, $F_{disk}/F_{star} = 8.2 \pm 0.8 \times 10^{-6}$ is about an order of magnitude (factor of 13) less than its thermal IR excess, suggestive, perhaps, of low albedo grains. We confirm through two-epoch non-common proper motions that three point-like objects seen initially in the Krist et al 2010 discovery imaging are not physically associated with the CDS, so if the ring is sculpted by one or more co-orbiting planets they remain undetected.

(3) For HD 202628, of comparable age to HD207129, we find its $a \approx$ 175 au debris ring more azimuthally isotropically scattering with $g = 0.1 \pm 0.1$ and confined to a sky-plane projected elliptical annulus from ~ 135 to 218 au. Its CDS morphology and structure is more complex with also: (a) a low-SB cloud of scattering material superimposed and exterior to the debris ring SE (only) ansa, (b) a stellocentric offset of the ring itself (initially suggested by Krist et al. 2012, confirmed here, but found in somewhat different amount), (c) a suggestion of spatially resolved debris "clumps" along the ring, most notably to the SW of the star. Together, (a) - (c) may implicate a much more dynamic environment perturbing the dust distribution absent in HD 202628. From (c) an $r^{-2}$ SB scaling would be expected with differing ansal distance, but is apparently modified by the presence of (a) at the SE ansa, further indication of a more asymmetrically complex overall dust density distribution. Qualitatively, (a) and (b) give the system when imaged at high sensitivity to large stellocentric distance (comparable to that of HD 207129) a distinctly asymmetrical "teardrop" morphology uniquely seen in HD 202628. Similar to HD 207129, though, HD 202628's $F_{disk}/F_{star} \approx 7.7. \times 10^{-6}$ is also an order of magnitude (factor of 18) less than its thermal IR excess.

(4) HD 202917 is the youngest (~ 30 My), by about two orders of magnitude, in age of the three close-solar analog CDSs in this sample. Its $a$ = 63 au, 13 au wide, debris ring is seen to be highly anisotropically scattering azimuthally. With $g$ = 0.6 ± 0.1, only the much brighter "half" of its 68.6° inclined debris ring, imaged through a scattering phase angle range ± 21°–159°, is detectable. HD 202917 possesses a highly complex and asymmetric exo-ring scattering structure, seen only beyond the brighter side of the ring itself and also with a strong exo-ansal brightness asymmetry. This "fan-like" exo-ring structure bears some morphological resemblance to the exo-ring "skirt" seen (at higher SNR) in the HD 61005 CDS suggestive, perhaps, of common origin or causation, though the anisotropy in ansal brightness may betray a recent conflating dust production event. Unlike HD 207129 and HD 202628, HD 201917's much brighter $F_{disk}/F_{star} \approx 3 \times 10^{-4}$ is nearly identical to its IR excess emission, as may be anticipated for younger CDSs; e.g., see Sch 14, *c.f.* Fig. 18.

(5) Combining the above three solar-analog CDSs with others we had previously observed as a small but homogeneous sample is cautiously suggestive of an $F_{disk}/F_{star}$ diminution of disk brightness over time (t) with a power law slope approximately $t^{-0.8}$. Though epochaly sporadic dust production/enhancement events in individual systems may bias this suggested trend (as exampled by HD 181327 and HD 61005).

(6) As an ensemble inclusive now of HD 15745, when imaged with sufficient depth and sensitivity as enabled with STIS 6R-PSFTSC, the number of CDSs exhibiting very large exo-ring scattering structures that are massive analogs to the outer regions of our solar system's Kuiper-belt is growing, and may be common. Such structures imaged in visible light uniquely trace the small grains that may be in the process of expulsion from these systems and/or interacting with their local interstellar environments (e.g., Fig. 14). Their presence inform that ongoing and future holistic models of CDS structure and evolution cannot treat debris systems in isolation.

Acknowledgements. Based on observations made with the NASA/ESA Hubble Space Telescope, obtained at the Space Telescope Science Institute (STScI), which is operated by the Association of Universities for Research in Astronomy, Inc., under NASA contract NAS 5-26555. These observations are associated with programs # 13786 and 12228. Support for program #13786 was provided by NASA through a grant from STScI. Carson acknowledges support from the South Carolina Space Grant Consortium.